\documentclass[10pt , a4paper]{article}
\usepackage[latin9]{inputenc}
\usepackage{amsmath,amsthm}
\usepackage{amsfonts}
\usepackage{amssymb}
\usepackage{graphicx}
\usepackage{bm}
\usepackage{amssymb}
\usepackage{authblk}
\usepackage{caption}

\providecommand{\keyword}[1]
{
  \small	
  \textbf{\textit{Keywords---}} #1
}

\begin{document}

\title{Congestion transition on random walks on graphs}


\author{Author Surname$^{1}$, Someone Else$^{2}$  \\
        \small $^{1}$University A \\
        \small $^{2}$University B \\
}

\author{Lorenzo Di Meco $^{1,2}$, Mirko Degli Esposti $^{1,2}$, Federico Bellisardi$^{1,2}$  and Armando Bazzani $^{1,2}$\\
        \small $^{1}$ \quad Department of Physics and Astronomy University of Bologna, Bologna, Italy; \\
        \small $^{2}$  \quad INFN sezione di Bologna, Bologna, Italy; \\
        e-mails: lorenzo.dimeco3@unibo.it, mirko.degliesposti@unibo.it, federico.bellisardi2@unibo.it, armando.bazzani@unibo.it
}
\maketitle





\abstract{The congestion formation on a urban road network is one of the key issue for the development of a sustainable mobility in the future smart cities.  In this work we propose a reductionist approach studying the stationary states of a simple transport model using of a random process on a graph, where each node represents a location and the weight links give the transition rates to move from one node to another that represent the mobility demand. Each node has a finite transport capacity and a maximum load capacity and we assume that the average. In the approximation of the single step process we are able to analytically characterize the traffic load distribution on the single nodes, using a local Maximum Entropy Principle. Our results explain how the congested nodes emerge when the total traffic load increases in analogous way to a percolation transition where the appearance of a congested node is a independent random event, However, using numerical simulations, we show that in the more realistic case of the synchronous dynamics for the nodes, there are entropic forces that introduce correlation among the node state and favor the clustering of the empty and congested nodes. Our aim is to highlight universal properties of the congestion formation and, in particular, to understand the role traffic load fluctuations as a possible precursor of congestion in a transport network.}

\keyword{Markov processes; Master equation; Entropic forces }

\section{Introduction}

Modeling city mobility is one of the main issues to plan future mobility infrastructures and to apply governance policies for a sustainable mobility in future smart cities\cite{batty2012,battybook}. A microscopic complex model to simulate urban mobility should define the individual mobility demand\cite{od1,od2}, simulate the individual decision mechanism to define the mobility strategies and to consider the effect of physical interactions on the transport networks\cite{barbosa2018}. This is a formidable task that requires an enormous amount of data to set up the model parameters and sophisticated methodologies to analyse the simulation results and to detect the control parameters of the system. Modeling  traffic dynamics has pointed out some universal features, like the stop and go congestion when the vehicle density overcomes a certain threshold\cite{nagatani2002}, that do not depend on the details of vehicle interaction. However, the emergence of congestion on an urban transport system is a different phenomenon\cite{barthelemy2011,manfredi2018} since the dynamics at the crossing points becomes more relevant than the dynamics on the roads. Previous papers have proposed the percolation theory as a key of lecture to understand the emergence of congestion on a traffic network\cite{li2014,wang2015,ambuhl2023}. Even the very definition of congestion on a road network can to be defined from different points of view: i.e. from a individual and from a macroscopic point of view. The individual perception is mainly affected by the change of average velocity of the mobility paths and the uncertainty on the travel time can be a fundamental information to understand the individual behavior performing urban mobility \cite{zheng2010,chen2019}. 
Conversely, the macroscopic approach considers the performance of the whole transport system introducing a macroscopic fundamental diagram\cite{daganzo2008} and modelling the network dynamics.
In this work we propose a reductionist approach to model a transport network dynamics and to study the congestion formation in stationary conditions by using of a random process on a graph where each node represents a location and the link weights (i.e. the transition rates
to move from one node to another) can be related to the statistical distribution on the mobility paths on the network\cite{bazzani2010} and to the individual mobility demand. Our aim is to highlight the universal features of congestion transition and the role of traffic load fluctuations. Indeed in a thermodynamic limit where both the number of nodes and of particles tend to infinity, but with a fixed ratio, the fluctuations remain finite and a average field approach is not suitable to describe the macroscopic evolution of the system. To highlight the effects of traffic fluctuations we explicitly study a balanced transport network where the average incoming and outgoing flows at each nodes are equal. This assumption could reflect the existence of a Wardrop equilibrium\cite{wardrop1952,dafermos1984} for urban traffic where the mobility paths distribute to optimize the use of the transport network. 
To simulate the traffic dynamics at crossing points we assume the existence of a finite transport capacity and of a maximal
capacity for each node so that a displacement is possible if the number of particles in the destination nodes is smaller than the maximal capacity\cite{helbing1,helbing2}.\par\noindent 
Our main assumption on the dynamics is that at a stationary state the traffic fluctuations can be modeled by a Markov process\cite{kindermann1980,ibe2013}. Indeed, we recall that a coarse grained description of a chaotic dynamical system allows to justify a stochastic approach. The Markov property means a short correlation in the traffic fluctuations among connected and it is probably justified if the mobility demand is spread on the urban fabric due to the city complexity\cite{battybook}.
\par\noindent
In the approximation of the single step process we are able to characterize the traffic load distribution of the nodes according to a local Maximum Entropy Principle\cite{jaynes1,jaynes2}. The application of the entropy concept to understand the statistical laws of urban mobility has been proposed in previous works\cite{gallotti2012,liang2013,mizzi2023} using a big-data approach. We extend this result to show how Entropy Principles allow to study the properties non-equilibrium states near a stationary state of the transport network model\cite{prigo1,auletta2017,endres2017}.
We are able to study as the congestion transition can be detected from the fluctuation statistics of the node traffic loads, whose variance has a maximum value when a peak at the congested nodes rises in the traffic load distribution, but there is no singularity in the thermodynamic limit. Using numerical simulations we show that the
macroscopic congestion in the network due to the emergence of a congested macroscopic cluster can be accurately explained by a percolation transition where the appearance of a congested node is an independent event. However, in the more realistic case of a synchronous dynamics for the transport network, we prove the appearance of entropic forces\cite{roos2014} that tend to cluster empty and congested nodes so that the size of the congested clusters increases even before the formation of a macroscopic congestion. The existence of small congested clusters can introduce a large variance on the travel time distribution for individual paths and our results suggest that their distribution can be used to characterize the congestion formation before the percolation transition.\par\noindent
The paper is organized as follows: in the second section we discuss how to use nonlinear stochastic Markov systems as dynamical models of a transport network near a stationary state: in the third section we illustrate the properties of random walks on graph as models of transport networks and we introduce an entropy based approach to study the traffic load fluctuations; in section four we study the congestion formation for a simple transport network model and. in the section five, we compare the analytical results and the numerical simulations. Finally we present some conclusions and perspectives.

\section{Methods: Modeling transport network as Markov processes}\label{sectransport}
Urban traffic is the consequence of the individual mobility demand to move from origins to destinations (OD) \cite{od1,od2}. However, the complex structure of modern cities\cite{batty2007} has ubiquitously distributed the activities in the urban fabric so that modelling human mobility by an approach based on the origin-destination paradigm is extremely difficult without collecting information at individual level \cite{barbosa2018}.  Moreover the realization of the mobility paths in the transport network\cite{barthelemy2011,manfredi2018} is the result of both of physical interactions (traffic dynamics) and of unpredictable individual decisions (free will) in the route choice. The statistical physics approach offer a possible solutions by using stochastic dynamical models related to coarse grained description of traffic dynamics, in particular when the macroscopic and mesoscopic states of traffic do not depend on the details of the individual dynamics. The unpredictable features of individual mobility demand justifies the use of random dynamical models to simulate the  mobility on transport networks and the application of maximum entropy principle (MEP) \cite{jaynes1,jaynes2} to study the properties of stationary states and the congestion transition. The existence of a \textit{mobility energy} (i.e. travel time budget) is consistent with the available data on individual mobility and it suggests that the mobility in a homogeneous transport network is characterized by an exponential path length distribution \cite{gallotti2012,liang2013}. The previous remarks suggest as a stochastic model for urban traffic can correctly described the statistical properties of the stationary states, but it could be inadequate to model the system out of equilibrium, when the complexity of urban mobility related to the individual behaviour could emerge.\par\noindent
The daily traffic load has strong periodic components and it is a reasonable assumption that each individual tries to optimize the daily mobility cost. One expects that the mobility paths distribute on the transport network to create a Wardrop equilibrium with respect the travel time cost. Moreover in normal conditions the mobility paths can be considered independent of the traffic load (i.e. individuals do not change their mobility strategies). As a consequence it is possible to study the path distribution and not the individual behavior to simulate the traffic dynamics. Moreover, the urban traffic is mainly dominated by the dynamics at crossing points which determine the maximal flow capacity \cite{helbing1}, while the path distribution defines the transition probabilities $\pi_{ij}$ to observe a vehicle passing from the $j$-road to the $i$-road. Using a reductionist point of view, we introduce a mathematical description of a transport network by using a graph where the nodes $i=1,...,M$ are either the roads or the crossing (our case), or sometimes bus/train stations or just generic transport facilities, and the links distribute the traffic load. Each node $i$ is characterized by an internal state $n_i$ (the traffic load) and the flow $\Phi_{ij}(n_i,n_j)$ between the node $j$ and the node $i$ depends on the node states (Markov field model\cite{kindermann1980}).
The existence of a maximal flow capacity implies that the total flow $\Phi_j$ outgoing from the $j$ node satisfies
$$
\sum_i \Phi_{ij}=\Phi_j\le \Phi^{\max}_j
$$
Let $\Delta t$ the evolution time scale if for a given the traffic load $n_j\le\Phi^{\max}\Delta t$ we have a free flow dynamics, conversely we have a queue formation at the node $j$, and the crossing time increases proportionally to the queue length (i.e. the traffic load $n_j$). The queue formation depends on the total traffic load on the transport network and one has a simple fundamental diagram  since the average velocity $\bar v$ decreases with the total traffic load $N$. When the load $n_j$ of a node approximates the maximal node capacity $n^{\max}$, the unavoidable traffic fluctuations can induce a gridlock and the incoming flows are drastically reduced. In this case the travel time may increase non-linearly with the total traffic load since it depends on how the congestion spreads in the network \cite{manfredi2018}. Assuming the existence of a stationary state of a transport network, the congestion formation is triggered by the traffic fluctuations.
If there are not long range spatial correlations in the mobility path distribution, the traffic dynamics at a stationary state can be approximated by a Markov random process whose transition probabilities $\pi_{ij}$ (i.e. a non-linear random walk on graph). The OD nature of the urban mobility is not relevant to define the stationary state when the correlation between the paths of two 'particles' in the same node is negligible (i.e. by randomly choosing two particles in the same node the probability that they have similar destinations is very small). Moreover, one can justify the relaxation process to a stationary state and to study the effect of fluctuations approximating the dynamics by a Markov process \cite{ibe2013,prigo1}. The average dynamics of the transport network reads
\begin{equation}
\dot n_i=\sum_j \left [\Phi_{ij}(n_i,n_j)-\Phi_{ji}(n_j,n_i)\right ]+s_i(t)
\label{average}
\end{equation}
where $s_i(t)$ represent the particle sources or the sinks present in the system that modulate the total traffic load. 
To study the stationary states we set $s_i=0$ and the traffic load $\sum_i n_i=N$ is constant. An average equilibrium solution satisfies
\begin{equation}
\sum_j \left [\Phi_{ij}(n_i^\ast,n_j^\ast)-\Phi_{ji}(n_j^\ast,n_i^\ast)\right ]=0\qquad \sum_j n_j^\ast=N
\label{eqstate}
\end{equation}
and the congestion transition occurs when the stationary solution becomes unstable and a new solution emerges with $n_j=n^{\max}$ for a subset of nodes (the congested nodes). When the congested nodes form a giant cluster in the transport network we say that the whole network is in a congested state.

\subsection{Characterizing the stationary state}
The equilibrium states are determined by the stationary flows among the connected nodes. By definition $\pi_{ij}$ is the probability that a particle performs the transition $j\to i$ and does not depend from the traffic load fluctuations. $\pi_{ij}$ is a stochastic matrix and in a free traffic condition one can set $\Phi_{ij}\propto \pi_{ij}n_j$.  However, a maximal transport capacity $\phi_i^{\max}$  and a maximal node capacity $n_i^{\max}$ exist (i.e. $\Phi_{ij}=0$ if $n_i\ge n_i^{\max}$). A possible definition for the flows $\Phi_{ij}(n_i,n_j)$ is
\begin{equation}
\Phi_{ij}(n_i,n_j)=\pi_{ij}\phi^{\max}_j c(n_i/n_i^{\max})\phi(n_j)
\label{flowdef}
\end{equation}
where the function $\phi(n_j)\in [0,1]$ is assumed monotonic increasing with an initial linear dependence and an asymptotic limit $\lim_{n\to\infty}\phi(n)=1$ (i.e. we do not consider a reduction of the outgoing flow when the node is congested assuming that the dynamics at crossing point is weakly affected by the road congestion), and the capacity function $c(n_i/n_i^{\max})\in [0,1]$ is threshold function that drops down to zero when $n_i\ge n_i^{\max}$. The flow and the capacity functions simulate the effect of particle interactions that affect the traffic dynamics. In the case of urban road network, the flow function $\phi(n)$ simulate by the traffic dynamics at crossing points. As a consequence of the definition (\ref{flowdef}), the transport network has a flow-density fundamental diagram when we increase the average traffic load for each node (cfr. Fig. \ref{fig:fdiagram} (left) in section \ref{sec_sim}). The equilibrium solution (\ref{eqstate}) follows from the condition
\begin{equation}
\sum_j \pi_{ij}\phi_j^{\max} c(n_i/n_i^{\max})\phi(n_j)=\sum_j \pi_{ji}\phi_i^{\max} c(n_j/n_j^{\max})\phi(n_i)
\label{eqstate1}
\end{equation}
If there exists a solution with $n_i\le n_i^{\max}$ the traffic load is sustainable, otherwise we have congested states with some nodes at the maximal capacity. In the case of a low traffic load, we expect $\phi(n_j)=\alpha n_j$ and $c(n_i/n_i^{\max})=1$ and eq. (\ref{eqstate1}) simplifies
$$
\sum_j \pi_{ij}\phi_j^{\max} n_j^\ast=\phi_i^{\max} n_i^\ast
$$
$n_i^\ast$ turns out to be the stationary eigenvector of the stochastic matrix $\pi_{ij}$ and the equilibrium state is stable since all the other eigenvalues $\lambda$ of the Laplacian matrix of the network
\begin{equation}
L_{ij}=\delta_{ij}\phi_i^{\max} -\pi_{ij}\phi_j^{\max}
\label{laplacian}
\end{equation}
have a negative real part if the network is connected \cite{vanBook}. When the traffic load increases the flow function $\phi(n)\to 1$ and we have to consider a self-consistent approach assuming $c(n_i/n_i^{\max})=1$ (i.e. no congestion in the network). We compute the null eigenvector $\phi_i^\ast$ 
$$
\sum_j \pi_{ij}\phi_j^{\max} \phi_j^\ast-\phi_i^{\max} \phi_i^\ast=0
$$
and we look for the solutions $n_i^\ast$ of the system
$$
\phi(n_i^\ast)\propto\phi_i^\ast\qquad \sum_i n_i^{\ast}=N
$$
We study the stability of the equilibrium solution in presence of perturbations $\delta n_j$ such that $\sum_j \delta n_j=0$ (i.e. the total traffic load is constant). The stability character depends on the derivative $\phi'=d\phi/dn$: if $\phi'(n_j^\ast)>0$ $\forall\; j$, then the linearized system
$$
\delta \dot n_i=-\sum_j L_{ij}\phi'(n_j^\ast)\delta n_j
$$
is still associated to a Laplacian matrix and the eigenvalues have all negative real part on the invariant subspace  $\sum_j \delta n_j=0$. When $\phi'(n_j^\ast)\to 0$ the system tends to a neutral stability. However, if the equilibrium traffic load $n_i^\ast$ increases up tp $n_i^{\max}$ the congested states appear due to the the congested function $c(n_i/n_i^{\max})$ in the definition (\ref{flowdef}). Assuming $n_i/n_i^{\max}\simeq	 1$ we can simplify eq. (\ref{eqstate1}) letting $\phi(n_j^\ast)\simeq \phi_j^{max}$ (i.e. before the congestion the roads express their maximum flow capacity), the equilibrium condition reads
$$
\sum_j \pi_{ij}\phi_j^{max}c(n_i/n_i^{\max})=\sum_j \pi_{ji}\phi_i^{\max}c(n_j/n_j^{\max})
$$
Therefore the solution has the form $c(n_i^\ast/n_i^{max})\propto c_i$ where $c_i$ is the null eigenvector of the Laplacian matrix
\begin{equation}
L'_{ij}=\sum_k \pi_{ik}\phi_k^{max}\delta_{ij}-\pi_{ji}\phi_i^{\max}  \label{laplacian2}
\end{equation}
(cfr. eq. (\ref{laplacian})) with the constraints
$$
\sum_i n_i^\ast=N\quad \textrm{and}\quad n_i^\ast<n_i^{\max}
$$
By increasing the traffic load $N$, $n_i^\ast\to n_i^{\max}$ for the nodes corresponding to smallest values of $c_i$ and the congestion will start from these nodes. 
To study the stability of the solution we consider the linearized system, i.e.:
$$
\delta \dot n_i=\sum_j L'_{ij} c'(n_j^\ast)\delta n_j\qquad c'(n_j^\ast)=\frac{dc}{dn_j}(n_j^\ast)
$$
and when $c'(n_j^\ast)< 0$ we have a stable congested state.\par\noindent
We observe that if $\Phi_j^{\max}$ is the stationary eigenvector of the stochastic matrix $\pi_{ij}$, we have $L_{ij}=L'_{ij}$ and the vectors $\phi_i^\ast$ and $c_i$ are constant (i.e. at the equilibrium states $\Phi_i/\phi_i^{\max}$ and $n_i/n_i^{\max}$ have the same value for all the nodes). A possible interpretation is that individuals use the transport network distributing the traffic load so that the nodes become equivalent from a dynamical point of view. This condition could reflect the emergence of a Wardrop equilibrium where the paths distribution is evolved to define a transition matrix $\pi_{ij}$ which makes all nodes equivalent in congestion formation: i.e. any change in the path distribution would make a node more susceptible to congestion.
Since the nodes are equivalent the congestion formation depends on the presence of traffic fluctuations that prevent the transport system to reach its maximal flow capacity and the congestion becomes a dynamical stationary state.
Indeed when a node $i$ is congested ($n_i\ge n_i^{\max}$), its incoming flow is null and some links $\pi_{ij}$ are ineffective.  Then the connected nodes $j$ have a greater probability to become congested, whereas the node $i$ can exit the congested state since the flow $\Phi_i>0$. We do not have an equilibrium state with fixed congested nodes, but a stationary dynamical state where congestion moves on the network in regions of almost congested nodes of the network. The rise of congestion due to traffic load fluctuations is not described by the average dynamics (\ref{average}) and the fluctuations define the traffic load stationary distribution on the transport network.

\section{Random walk on graphs as models of transport networks}
To study the physics of traffic fluctuations we propose simple models based on random walks on graph \cite{vanBook}.  According to the previous section 
we associate a graph to a transport network where the transition probabilities $\pi_{ij}$ define the link weights.
In this way we can define a Markov process where the 'particles' move randomly according to the transition probabilities. In this way it is possible to study the fluctuations statistics at stationary states and the congestion formation.  
To perform an analytical approach  we simplify the model (\ref{flowdef}) by defining the flow function $\phi(n)=\Theta(n)$ where $\Theta(n)$ is the Heaviside function (i.e. $\Theta(n)=1$ if $n\ge 1$ and $\Theta(0)=0$) and the capacity function $c(n)=\Theta(n^{\max}-n)$ where all the nodes have the same maximum capacity $n_{max}$. We refer to the vector $\vec n$ whose components $n_i\ge 0$ $i=1,...,M$ give the number of particles at the node $i$, as the dynamical state of the network and we define $\vert \vec n\vert=\sum_i n_i=N$ the traffic load of the transport network.
In the one step process approximation, the evolution of the distribution function $\rho(\vec n,t)$ is given by the master equation
\begin{equation}
\frac{1}{M}\dot \rho(\vec n,t)=\sum_{(i,j)}\left [E_i^- E_j^+ \Phi_{ij}(n_i,n_j)-\Phi_{ji}(n_j, n_i)\right ]\rho(\vec n, t)
\label{master1}
\end{equation}
where we set the transition rates
\begin{equation}
\Phi_{ij}(n_i,n_j)=\Theta(n_j)\Theta(n^{\max}-n_i)\pi_{ij}
\label{weight}
\end{equation} 
and the sum runs over all the possible ordered couples $(i,j)$. We remark that the one step process implies an instantaneous upgrade of the information after each movement. This is a non-physical assumption for a transport network, and we also consider the synchronous dynamics where all the nodes evolve simultaneously and the information of the network state is upgraded after the movement of all the nodes. In this case the master equation becomes more complicated and an analytical approach is not possible. In the Appendix \ref{append1} we show how to write the incoming and outgoing flows for a generic node (see eq. (\ref{flowsync})).\par\noindent 
We observe that if $j\ne i$ we have the identity
$$
E_i^- E_j^+ \pi_{ij}\Theta(n_j)\Theta(n^{\max}-n_i)\rho(\vec n)=\Theta(n_i)\Theta(n^{\max}-n_j)\pi_{ij}\rho(\vec n+\hat e_j-\hat e_i)
$$
where we set $\rho(\vec n)=0$ for any non-physical state and we write the master equation in the form
\begin{equation}
\frac{1}{M}\dot \rho(\vec n,t)=\sum_{(i,j)}  \Theta(n_i)\Theta(n^{\max}-n_j)\left [\pi_{ij}\rho(\vec n+\hat e_j-\hat e_i,t)-\pi_{ji}\rho(\vec n, t)\right ]
\label{masterm}
\end{equation}

\subsection{Equilibrium state in the case of detailed balance}
We are interested in the stationary distribution $\rho_s(\vec n)$ of the master equation (\ref{masterm}). 
Let $\vec p$ the null right eigenvector of the Laplacian matrix of the weighted network (i.e. $p_i$ is the probability to observe a particle in the node $i$ according to the transition rates (\ref{weight}))
\begin{equation}
\sum_j \pi_{ji}p_i -\pi_{ij} p_j=0\qquad \sum_j p_j=1 \quad p_j>0
\label{laplacian1}
\end{equation}
we look for a stationary distribution of the form
\begin{equation}
\rho_s(\vec n\,)=\begin{cases}[C_N^{\, n^{\max}}]^{-1}(\vec p\,)\prod_{k=1}^M p_k^{n_k} & \qquad |\vec n|=N\quad n^{\max}\ge n_i\ge 0\\
 0 & \mbox{otherwise }  \end{cases}
\label{distri1}
\end{equation}
$C_N^{n^\ast}(\vec p\,)$ is the normalizing constant (i.e. the partition function) and it is related to the Helmholtz Free Energy
\begin{equation}
F(\vec p\,)=-\ln C_N^{\, n^{\max}}(\vec p\, )=-\ln \sum_{|\vec n|=N}^{n^{\max}} \prod_{i=1}^M p_i^{n_i}
\label{partitionf}
\end{equation}
where the sum runs over the physical states $n_i\le n^{\max}$ 
(see eq. (\ref{cncomp}) in the Appendix \ref{append1} for some analytical estimates in the case $n^{\max}\gg 1$).
The stationary condition reads
\begin{equation}
\sum_{(i,j)}\Theta(n_i)\Theta(n^\ast-n_j) \left [\pi_{ij} \frac{p_j}{p_i}-\pi_{ji}\right ]\rho_s(\vec n)=0
\label{stazeq}
\end{equation}
that has to be satisfied for all the physical states $\vec n$. The main difficulty to get an analytical solution of eq. (\ref{stazeq}) is the presence of empty and congested nodes at the same time. However, if the micro-dynamics  satisfies a detailed balance (DB) condition
\begin{equation}
\pi_{ij}p_j=\pi_{ji}p_i
\label{db}
\end{equation}
the distribution (\ref{distri1}) is the stationary solution of the master equation (\ref{masterm}). 
The DB condition means that in the stationary state the probability to observe a path between the nodes $j\to i$ is the same if we consider the reverse displacement $i\to j$. We remark that the stationary distribution $\vec p$ depends on the transition matrix $\pi_{ij}$, but different transition matrices have the same distribution. The DB condition associates uniquely the transition matrix to the distribution $\vec p$. If we introduce the Gibbs Entropy
\begin{equation}
\mathcal{S}[\rho(\vec n\,)]=-\sum_{|\vec n|=N}^{n^{\max}}\rho(\vec n\,)\ln\rho(\vec n\,)
\label{gibbsent}
\end{equation}
the DB condition allows to apply a Maximum Entropy Principle (MPE) to computed the stationary distribution.
Let $\bar n_i$ are the mean load per node
\begin{equation}
\bar n_i=\sum_{|\vec n|=N}^{n^{\max}} n_i\rho_s(\vec n\,)\qquad i=1,...,M
\label{constraint}
\end{equation}
the Gibbs entropy (\ref{gibbsent}) is maximal for the distribution (\ref{distri1}) when the probabilities $p_i$ satisfy the constraints (\ref{constraint}). The thermodynamic approach allows to characterize the statistical properties of the stationary state without considering the dynamics (\ref{master1}). An explicitly computation of the entropy for a distribution of the form (\ref{distri1}) gives
$$
S(\vec p\,, N)=\ln C_N^{\, n^{\max}}(\vec p\,)-\sum_{|\vec n|=N}^{n^{\max}} \rho_s(\vec n\,) \sum_i n_i \ln p_i=-F(\vec p\,)-\sum_i \bar n_i \ln p_i
$$
and we get
\begin{equation}
\frac{\partial S}{\partial p_i}=-\frac{\partial}{\partial p_i}F(\vec p\,)-\frac{\bar n_i}{p_i}
\label{aveval}
\end{equation}
The extremal condition reads
$$
\bar n_i=-p_i \frac{\partial}{\partial p_i}F(\vec p\,)
$$
and it is verified by the distribution (\ref{distri1}). 
When one increases the total traffic load $N$, we observe that the distribution (\ref{distri1}) is peaked on the marginal states where the nodes with the greater $p_i$ are congested $n_i=n^{\max}$ and the other nodes are empty. This means that the nodes with greater $p_i$ are hot spots for the congestion spreads and the mitigation policies have to reduce the traffic load on these nodes by redistributing the mobility paths. The situation is different when one consider an homogeneous transport network where $p_i=M^{-1}$ for all the nodes. Among all the possible distributions of the form (\ref{distri1}) this provides the maximum value for the Gibbs entropy and the congestion formation is triggered by the traffic fluctuations since the $\bar n_i=N/M$ for any node.
\par\noindent
To study the relation between entropy and the traffic load fluctuations, we compute the Hessian matrix of the entropy with respect the variation of $p_i$
\begin{align}
\label{entropyfluct}
&p_j p_i\frac{\partial^2 S}{\partial p_j\partial p_i}=p_j p_i\frac{\partial}{\partial p_j}p_i^{-1}\left (-p_i\frac{\partial}{\partial p_i}F(\vec p)-\bar n_i\right )=\nonumber \\
&=-p_j\frac{\partial}{\partial p_j}[C_N^{\, n^\ast}]^{-1}\sum_{|\vec n|=N}^{n^{\max}} n_i\prod_{k=1}^M p_k^{n_k}=-\left (\left\langle n_i n_j \right \rangle-\bar n_j \bar n_i\right )
\end{align}
where we have used eq. (\ref{aveval}). When the system is in a stationary states the covariance matrix of the traffic loads $n_i$ is the sensitivity of the entropy to the changes of the probability distribution $p_i$.  Since the distribution $p_i$ depends on the transition probabilities $\pi_{ij}$, the previous equation gives also a measure of the sensitivity to the perturbations of the transition matrix. Using the entropy as a measure of the disorder in the system, the result (\ref{entropyfluct}) means that when the fluctuations of the traffic load $n_i(t)$ are large, the system may pass through ordered and disordered states during the evolution. We also have the reciprocity relations that holds in the DB condition
$$
p_j\frac{\partial \bar n_i}{\partial p_j}=p_i \frac{\partial \bar n_j}{\partial p_i}=-p_i p_j\frac{\partial F}{\partial p_i\partial p_j}\qquad i\ne j
$$
to define how the average load of different nodes is affected by the change of the stationary probabilities: these relations correspond to the Onsager reciprocal relations.\par\noindent 
In the case of a homogeneous network the fluctuation variance at each node is a key indicator for understanding the rise of congestion.
One expects that the fluctuations reduce when the traffic load is low ($n_i\ll n^{\max}$) or highly congested ($n_i\simeq n^{\max}$), so that one get a critical value for the total traffic load when the average fluctuation variance is maximal. Using the covariance matrix with $Np_i=n_i$
$$
\sigma_{ij}(N)=-n_in_j \frac{\partial^2 S}{\partial n_j\partial n_i}(N)
$$
we introduce the following definition
$$
\bar \sigma(N)=\frac{1}{M}\textrm{Tr}\, \sigma=\frac{1}{M}\sum_i n_i^2 \frac{\partial^2 S}{\partial n_i^2}(N)
$$
and the critical load  $N_c$ satisfies 
\begin{equation}
\frac{\partial \bar\sigma}{\partial N}(N_c)=0
\label{critval}
\end{equation}
Using numerical simulations on a simple traffic network model we have computed the traffic load variance for the traffic load of single nodes as a function of the average traffic load (see Fig. \ref{fig:fdiagram} (right) in section \ref{sec_sim}). The simulations show that the maximum value of the standard deviation is achieved at the critical value of the flow-density fundamental diagram . The critical load $N_c$ corresponds both to the maximum flow in the transport network and to the maximum uncertainty in traffic load distribution. When $N\ge N_c$ the congested nodes start to merge in clusters until a macroscopic large cluster emerges in the network.
The congestion degree can be related to the dimension of congested clusters that is the fingerprint of a percolation phase transition for the congestion formation\cite{ambuhl2023}.

\subsection{Non-equilibrium stationary states}
When the detailed balance condition (\ref{db}) does not hold, the Markov process to model the vehicle dynamics realizes a non reversible random walk: i.e. the statistics of the reverse paths on the transport network is not equivalent to the statistics of the original paths in the stationary states. As a consequence we have the presence of probability stationary currents on the links $j\to i$ defined by
\begin{equation}
J_{ij}(\vec n)=\Theta(n_i)\Theta(n^{\max}-n_j)\left [\pi_{ij}\rho_s(\vec n+\hat e_j-\hat e_i,t)-\pi_{ji}\rho_s(\vec n, t)\right ] 
\label{currents}
\end{equation}
that correspond to net traffic flows moving on the loops of the transport network. In such a case the MEP cannot be applied \cite{auletta2017}, but it is possible to maximize a local entropy or applying an entropy production principle \cite{endres2017}. If we consider the ensemble of states $\vec n$ with $0\leq n_j < n^{\max}$ so that $\Theta(n^{\max}-n_j)=1$, then the stationary solution satisfies
$$
\sum_j \left [\pi_{ij}\rho_s(\vec n+\hat e_j-\hat e_i,t)-\pi_{ji}\rho_s(\vec n, t)\right ] =0
$$ 
and we recover a solution of the form (\ref{distri1}) which maximizes the Entropy if restricted to this ensemble (cfr. eq. (\ref{stazeq})). Conversely if one considers the ensemble of states  $\vec n$ with $0<n_i \le n^{\max}$ and $\Theta(n_i)=1$ and we get the condition
$$
\sum_i \left [\pi_{ij}\rho_s(\vec n+\hat e_j-\hat e_i,t)-\pi_{ji}\rho_s(\vec n, t)\right ] =0
$$ 
and we have a solution of the form
\begin{equation}
\rho_s(\vec n\,)\propto \prod_{k=1}^M q_k^{-n_k}
\label{distri2}
\end{equation}
where $\vec q$ is the stationary eigenvector of the Laplacian matrix associated to the reverse transition matrix $\pi_{ij}^T$. The solution (\ref{distri2}) approximates the stationary distribution in the case of high traffic loads and it is a maximum entropy solution for the congested states.
The distribution (\ref{distri2}) is the distribution for the gaps dynamics on a congested transport network. In the case of DB $q_i=p_i^{-1}$ and the two distributions coincide. Changing the traffic load the stationary distribution $\rho_s(\vec n)$ changes its forms interpolating the two limit distributions. The congestion transition depends probability of intermediate states, where one has both empty and congested nodes.

\section{Results: congestion formation in balanced transport networks}
\label{congsec}
To understand the role of traffic fluctuations in the congestion transition we consider a homogeneous transport network where all the roads are equivalent $p_i=const.$. Using the simple model (\ref{weight}) this condition corresponds to the balance condition 
\begin{equation}
\sum_i \pi_{ij}=\sum_j \pi_{ij}
\label{balance1}
\end{equation}
so that for each node the expected incoming flow is balanced by the expected outgoing flow. This condition is consistent with the existence of a Wardrop's equilibrium in the transport systems. We also remark that the condition (\ref{balance1}) does not imply the detailed balance that requires the symmetry of the transition rates $\pi_{ij}$. The stationary solution can be approximated by a uniform distribution (cfr. eq. (\ref{distri1})) for the majority of the microstates, but not for the network states $\vec n$ in which they are simultaneously present empty and congested connected nodes. Due to the traffic fluctuations, these states are more probable when the average traffic load is $N/M=n^{\max}/2$ where probability of the empty and the congested nodes equals, so that they can play a role in the congestion transition.\par\noindent
To study how the node states are distributed in the network, we consider the single node distribution $p(n)$ defined as the probability to observe a node in the state $n$ in the stationary state. In an homogeneous transport network all the node are equivalent so that the marginal distribution for a single node of the stationary distribution $\rho_s(\vec n)$ approximates the traffic load distribution on the network in the thermodynamic limit.
The distribution $p(n)$ can be related to the congestion degree 
by computing the probability of the almost congested states
$$
C(\tau,N)\simeq \sum_{n\ge n^{\max}-n(\tau)} p(n)
$$
where a suitable $n(\tau)$ is a decreasing function of $\tau$.\par\noindent
In the case of low traffic load $N/M\ll n^{\max}$ it is possible to prove that in the thermodynamics limit the single node distribution $p(n)$ is approximated by an exponential distribution
\begin{equation}
p(n)\simeq \frac{1}{\bar n}\exp\left ( -\frac{n}{\bar n}\right )\qquad \bar n=\frac{N}{M}
\label{expapp0}
\end{equation}
and the congestion probability is exponentially small. Analogously, in the case of high traffic load (i.e. $N\simeq n^{\max}M$) is approximated by
\begin{equation}
p(n)\propto \exp\left ( -\frac{n^{\max}-n}{n^{\max}-\bar n}\right )\qquad n\le n^{\max}
\label{expapp1}
\end{equation}
(see Appendix \ref{append1} for the proof) that explains as the number of congested nodes $C(\tau,N)$ increases with the traffic load $N$.
The true distribution interpolates the two approximations as the traffic load varies, but this is a continuous process and the distribution does not cross any singularity (see section \ref{sec_sim}).
\par\noindent
In the case of a synchronous dynamics, each node moves one particle, but can receive a variable number of particles depending on the connectivity degree. An analytic expression for the stationary distribution $\rho_s(\vec n)$ is not available and we remark that the synchronous dynamics introduces a correlation among the states of connected nodes.  Indeed a node which is connected with empty nodes has a reduced incoming flow in average, and its population tends to reduce. As a consequence, the network states where the empty nodes cluster are favored by the synchronized dynamics. Analogously, if a node is connected to congested node its outgoing flows reduces in average and its population increases so that the congested nodes tend to cluster too. One can interpret this fact introducing entropic forces that attract the empty and the congested nodes in the synchronous dynamics. In the last part of Appendix \ref{append1} we discuss the relation between the synchronous dynamics and the appearance of the entropic forces.  In the section \ref{sec_sim} we study the effect of entropic forces by using numerical simulations.

\subsection{Single node dynamics}
We now consider the following problem: 
\begin{quote}
    is it possible to introduce an effective model for the node dynamics to quantify the effect of entropic forces?

\end{quote}
The node state evolution can be modeled by an effective master equation that depends on the neighbor node states. In the one-step process model it is possible to look for a self-consistent master equation for the distribution $p(n)$.  
One can consider the balance between the incoming and outgoing flows of a representative node $i$ given its state $n\le n^{\max}$ average over all the states of the network. The incoming flow is the transition probability from the state $n-1\to n$ so that $n\ge 1$ and we get
$$
\left \langle \sum_j \pi_{ij}\Theta(n_j)-\sum_j \pi_{ji}\Theta(n^{\max}-n_j)\right \rangle=0
$$
One applies the following estimates
$$
\langle \Theta(n_j) \rangle=1-\pi(0|n-1)p(n-1)\qquad \langle\Theta(n^{\max}-n_j)\rangle=1-\pi(n^{\max}|n)p(n)
$$
where $\pi(0|n-1)$ is the conditional probability that a neighbor is empty when the node state is $n_i=n-1$ and $\pi(n^{\max}|n)$ s the conditional probability that a neighbor is congested when the node state $n_i=n$. Using the condition (\ref{balance1}) we have the equilibrium 
\begin{equation}
(1-\pi(0|n-1))p(n-1)-(1-\pi(n^{\max}|n)) p(n)=0 \qquad 0<n\le n^{\max}
\label{stazsinglenode}
\end{equation}
The conditional probabilities may depend on the degree of the node: i.e.  and the average  is the result of an averaging on all the nodes.  Eq. (\ref{stazsinglenode}) can be solved recursively using $p(0)$ to normalize the distribution.
We remark that the conditional probabilities are estimated from the global dynamical properties of the transport network and measure the correlation among the states of connected nodes.
In the one step process case the distribution is derived from a MEP and we expect that the node states are independent so that in a thermodynamic limit $\pi(0|n)=p(0)$ and $\pi(n^{\max}|n)=p(n^{\max})$. 
In the synchronous dynamics the stationary distribution does not maximize the entropy and we expect that $\pi(0|n)$ is a decreasing function on $n$ since the empty nodes tend to cluster and $\pi(n^{\max}| n)$ is an increasing function. These effects are consequence of the stochastic dynamics when the network state contains empty and congested nodes at the same time and it creates entropic forces that explains the correlation among the connected nodes.

\section{Results: numerical simulations of the transport network model}
\label{sec_sim}
We have checked the applicability of the analytical by simulating the simplified model (\ref{weight}) using a random network of $500$ nodes with average degree $d=3$ (but the minimum degree is $d=2$). For a given transition matrix $\pi_{ij}$ we have computed the stationary solution $p_i$ and we have define the stationary flows $\Phi_{ij}=\pi_{ij}p_j$ assuming $\phi^{\max}=1$, we have the balance condition
$$
\sum_j \Phi_{i j}=\sum_j \pi_{i j}p_j=p_i=\sum_j \pi_{j i}p_i=\sum_i \Phi_{i j}
$$
As previously discussed this condition means that the flows distribute on the transport network so that the average incoming flows  equal the average outgoing flows for each node simulating an optimal use of the network. The maximal average flow for the whole network is defined by
\begin{equation}
    \bar \Phi^{\max}=\frac{1}{M}\sum_{ij} \Phi_{ij}
    \label{aveflowmax}
\end{equation}
and the maximal node capacity is fixed at $n^{\max}=10$. In the sequel we refer to this model as the \textit{transport network model}. The balance condition (\ref{balance1}) for the flows highlight the role of fluctuations in the emergence of the congested states, but we remark that the transition matrix $\pi_{ij}$ does not satisfy the DB condition (i.e. $\Phi_{ij}$ is not symmetric). This model is used in all the simulations presented in the sequel.\par
\begin{figure}
    \begin{minipage}[t]{0.5\textwidth}
        \centering
        \includegraphics[width=\linewidth]{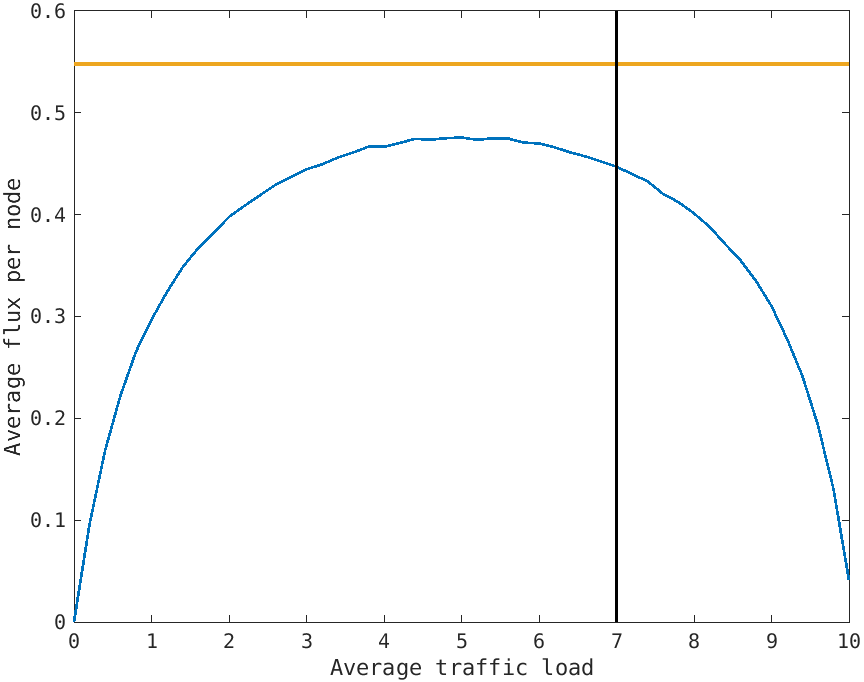} 
    \end{minipage}
    \hfill
    \begin{minipage}[t]{0.5\textwidth}
        \centering
        \includegraphics[width=\linewidth]{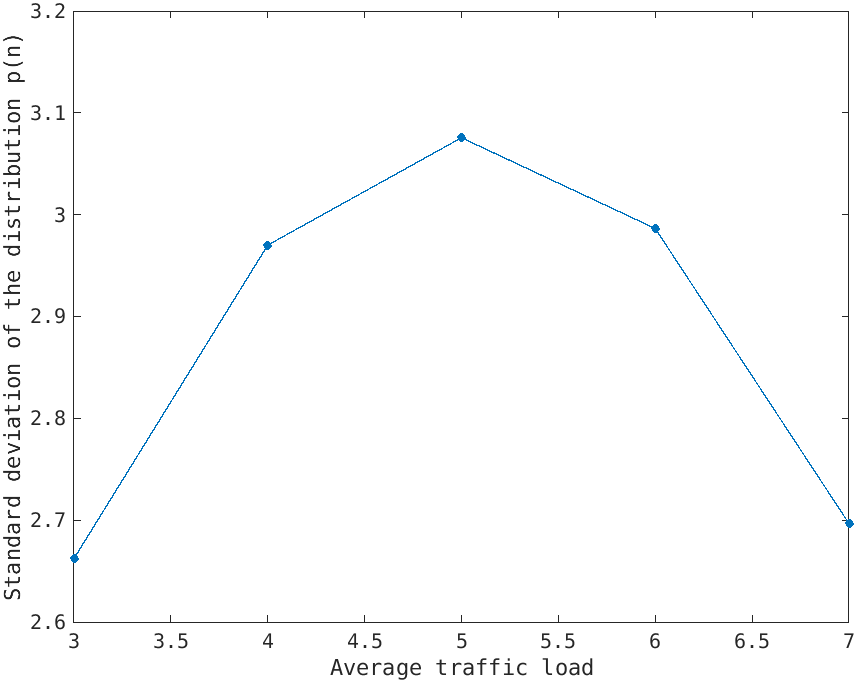} 
    \end{minipage}
\caption{Left picture: fundamental diagram for the transport network model using the synchronous dynamics where the average flow in plotted as a function of the average traffic load. The horizontal line denotes the maximal possible average flow in the network that is greater than the numerical value achieved at $\bar n=n^{\max}/2=5$ as a consequence of the presence of empty and congested nodes at any traffic load due to traffic fluctuations. The vertical line at $\bar n=7$ is the critical traffic load at which the small congested clusters start to merge (see Fig. \ref{fig:percolation}). Right picture: standard deviation of the single node traffic load distribution using the transport model in the synchronous dynamics as a function of the average traffic load $3\le \bar n\le 7$. The maximum value of the standard deviation is obtained critical value $\bar n=5$ of the traffic load.}
\label{fig:fdiagram}
\end{figure}

\begin{figure}
    \begin{minipage}[t]{0.5\textwidth}
        \centering
        \includegraphics[width=\linewidth]{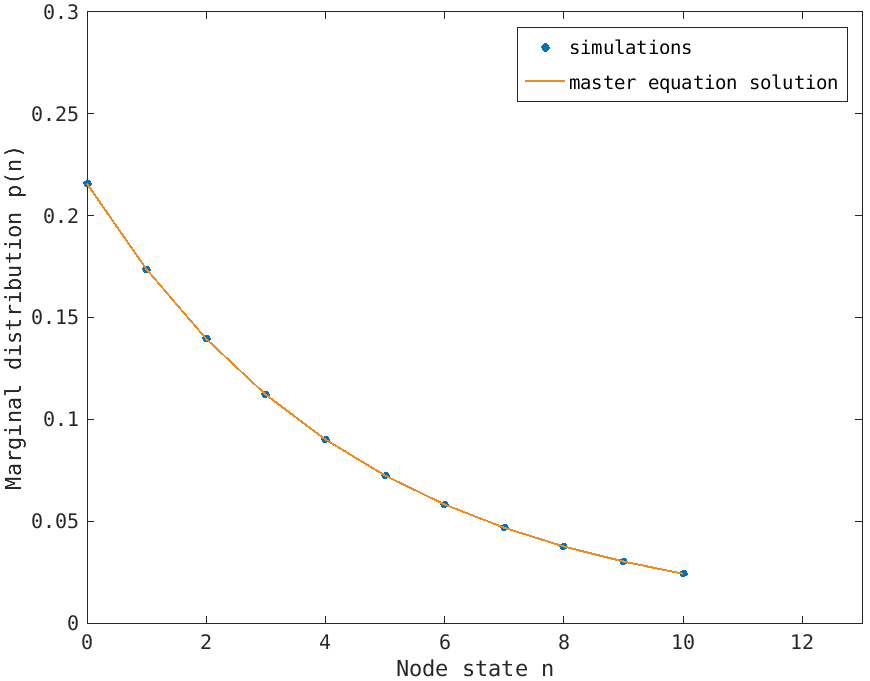} 
        \caption*{One-step dynamics with $\bar n=3$.}
    \end{minipage}
    \hfill
    \begin{minipage}[t]{0.5\textwidth}
        \centering
        \includegraphics[width=\linewidth]{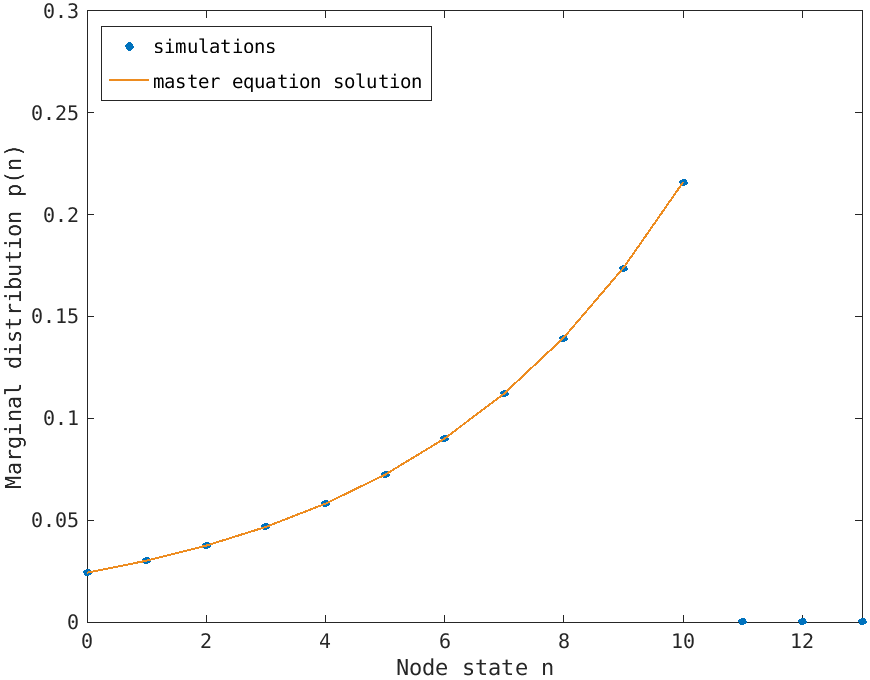} 
        \caption*{One-step dynamics with $\bar n=7$.}
    \end{minipage}
    \vskip\baselineskip
    \begin{minipage}[t]{0.5\textwidth}
        \centering
        \captionsetup{justification=centering}
        \includegraphics[width=\linewidth]{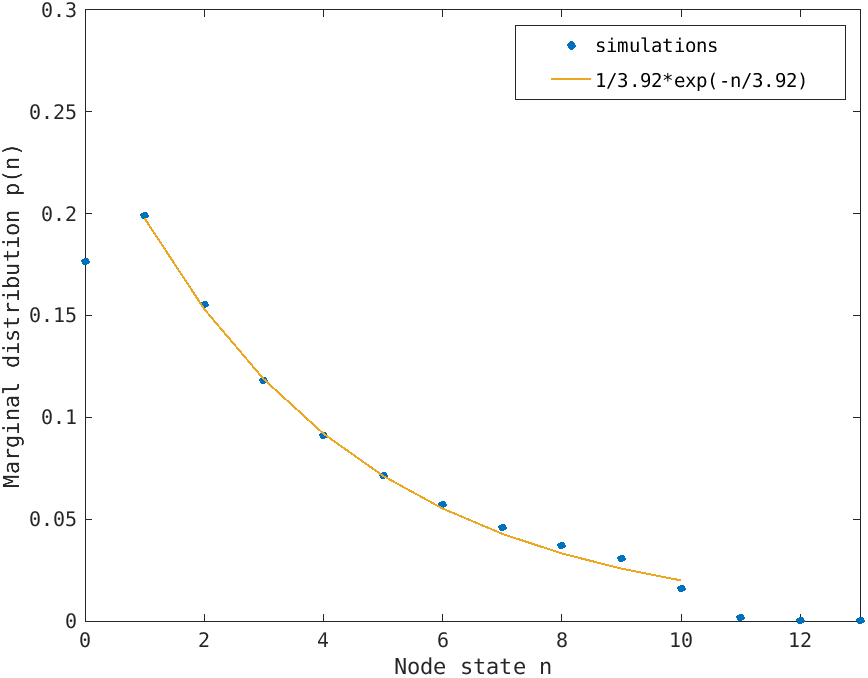}
        \caption*{Synchronous dynamics with $\bar n=3$.}
    \end{minipage}
    \hfill
    \begin{minipage}[t]{0.5\textwidth}
        \centering
        \captionsetup{justification=centering}
        \includegraphics[width=\linewidth]{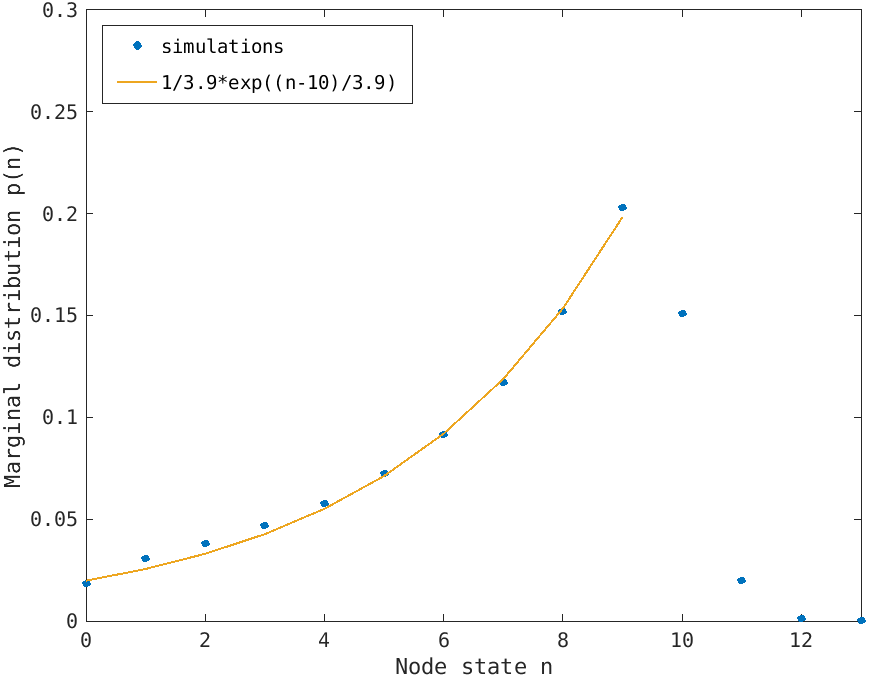}
        \caption*{Synchronous dynamics with $\bar n=7$.}
    \end{minipage}
    \caption{Top pictures: the dots represent the empirical distributions $p(n)$ of the node traffic load for the  transport network model in the one step process approximation. The distributions are computed by averaging over $10^5$ evolution steps of the model. The maximum node capacity is $n^{\max}=10$ and we consider two average traffic loads: a low traffic load $\bar n=3$ (left picture) and a high traffic load $\bar n=7$ (right picture). The continuous lines refer to the stationary solution of the single node dynamics (\ref{stazsinglenode}).   
    Bottom pictures: the dots represent the empirical distributions of the node traffic load for the  transport network model using the synchronous dynamics. The parameters used in the simulations are the same as in the top pictures. The continuous line is an exponential interpolation according to eqs. (\ref{expapp0}) and (\ref{expapp1}).}
    \label{fig:p_n_3_and_7}
\end{figure}

\begin{figure}
    \centering
    \begin{minipage}[t]{0.58\textwidth}
        \centering
        \includegraphics[width=\linewidth]{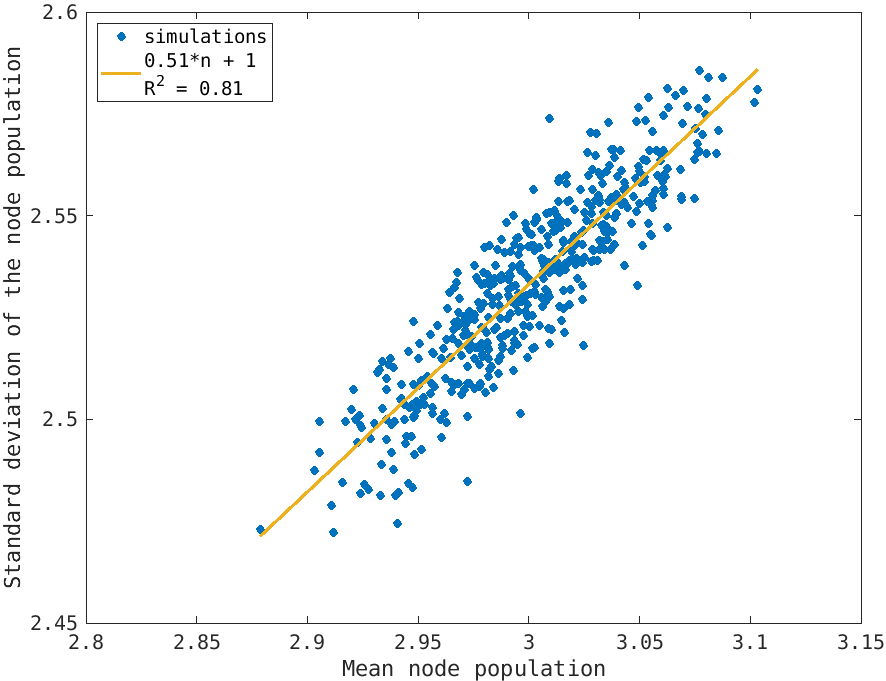}
        \label{fig:sub_1}
    \end{minipage}
    \vskip\baselineskip
    \begin{minipage}[t]{0.58\textwidth}
        \centering
        \includegraphics[width=\linewidth]{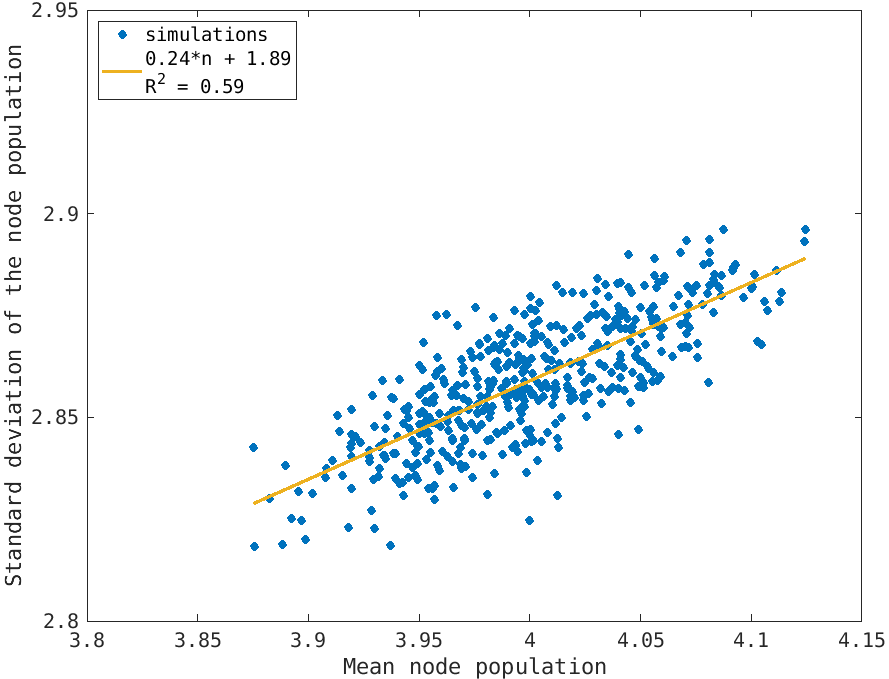}
        \label{fig:sub_2}
    \end{minipage}
    \vskip\baselineskip
    \begin{minipage}[t]{0.58\textwidth}
        \centering
        \includegraphics[width=\linewidth]{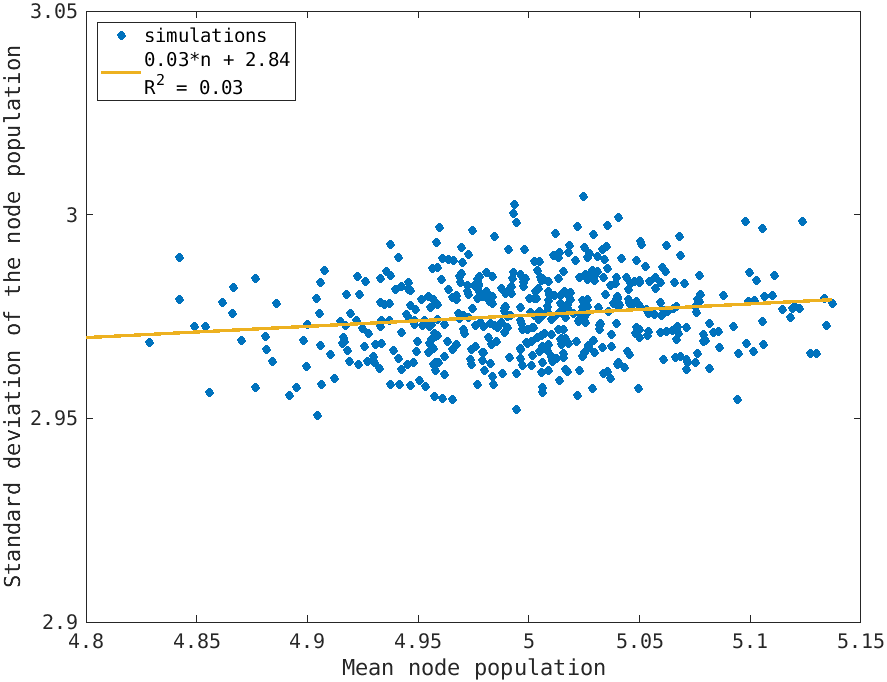} 
        \label{fig:sub_3}
    \end{minipage}
    \caption{Standard deviation of the single node traffic load for the transport network model in the synchronous dynamics case as function of the corresponding mean value. From top to bottom the figures refer to an average traffic load per node $\bar n= 3,4,5$ (from top to down) and the continuous lines refer to the linear interpolation whose parameters are reported in the insets.}
    \label{fig:std_devs_mean}
\end{figure}

\begin{figure}
    \begin{minipage}[t]{.5\textwidth} 
    \centering
    \includegraphics[width=1\textwidth]{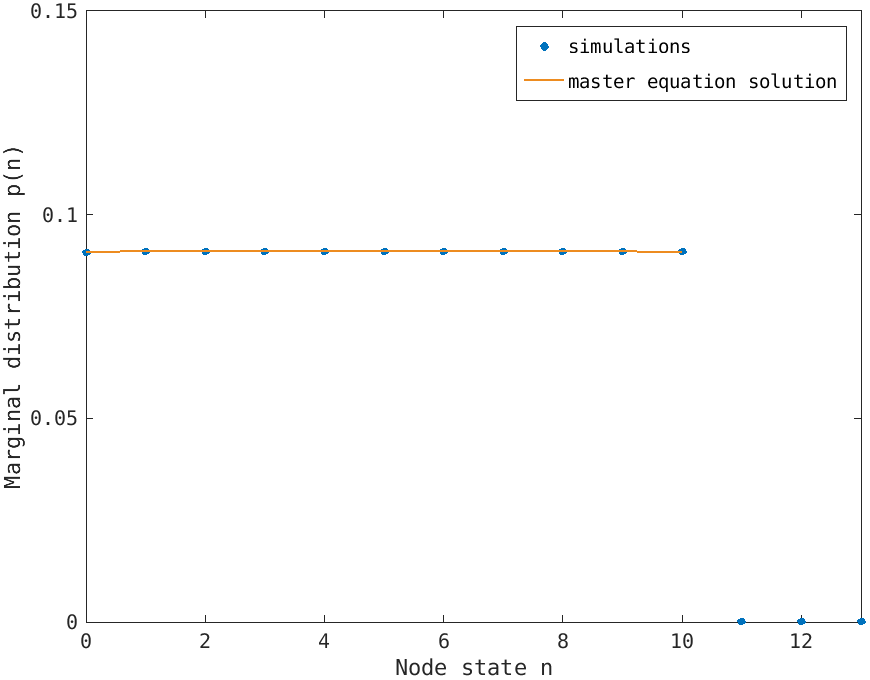} 
    \end{minipage}
\hfill
    \begin{minipage}[t]{.5\textwidth}
    \centering
    \includegraphics[width=1\textwidth]{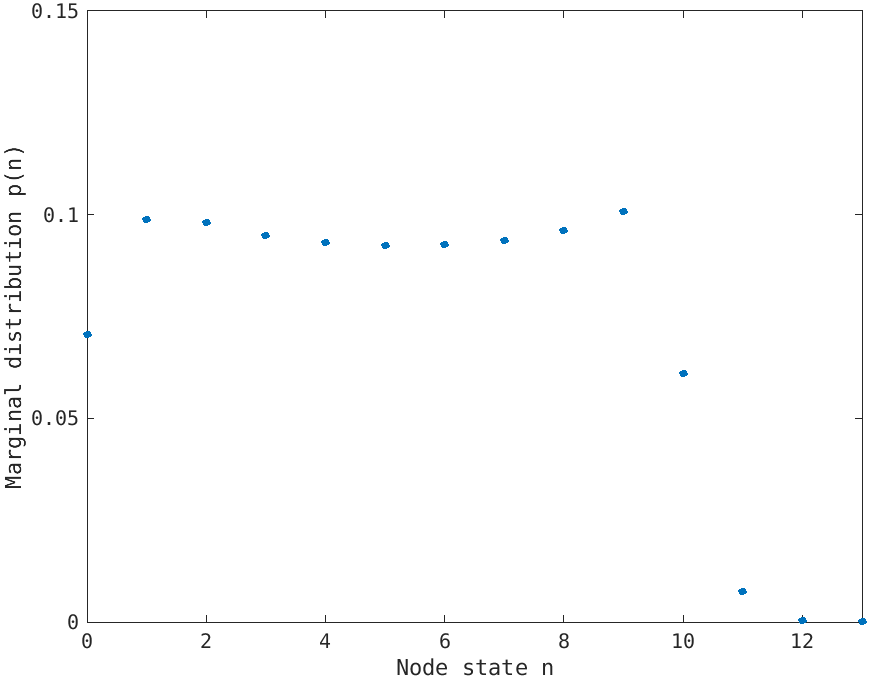}
    \end{minipage}
\caption{Stationary distribution probability for the transport network model using the same parameters as in Fig. \ref{fig:p_n_3_and_7} at the critical value of the traffic load $\bar n=5$. The left picture refers to the one-step process and the empirical distribution is almost flat (it's actually slightly peaked $n=5$), whereas the right picture refers to the synchronous dynamics and the distribution is bimodal near the empty and congested states, but we have not an analytical approximation.}
\label{fig:p_n_5}
\end{figure}

\begin{figure}
    \begin{minipage}[t]{.49\textwidth} 
    \centering
    \includegraphics[width=1\textwidth]{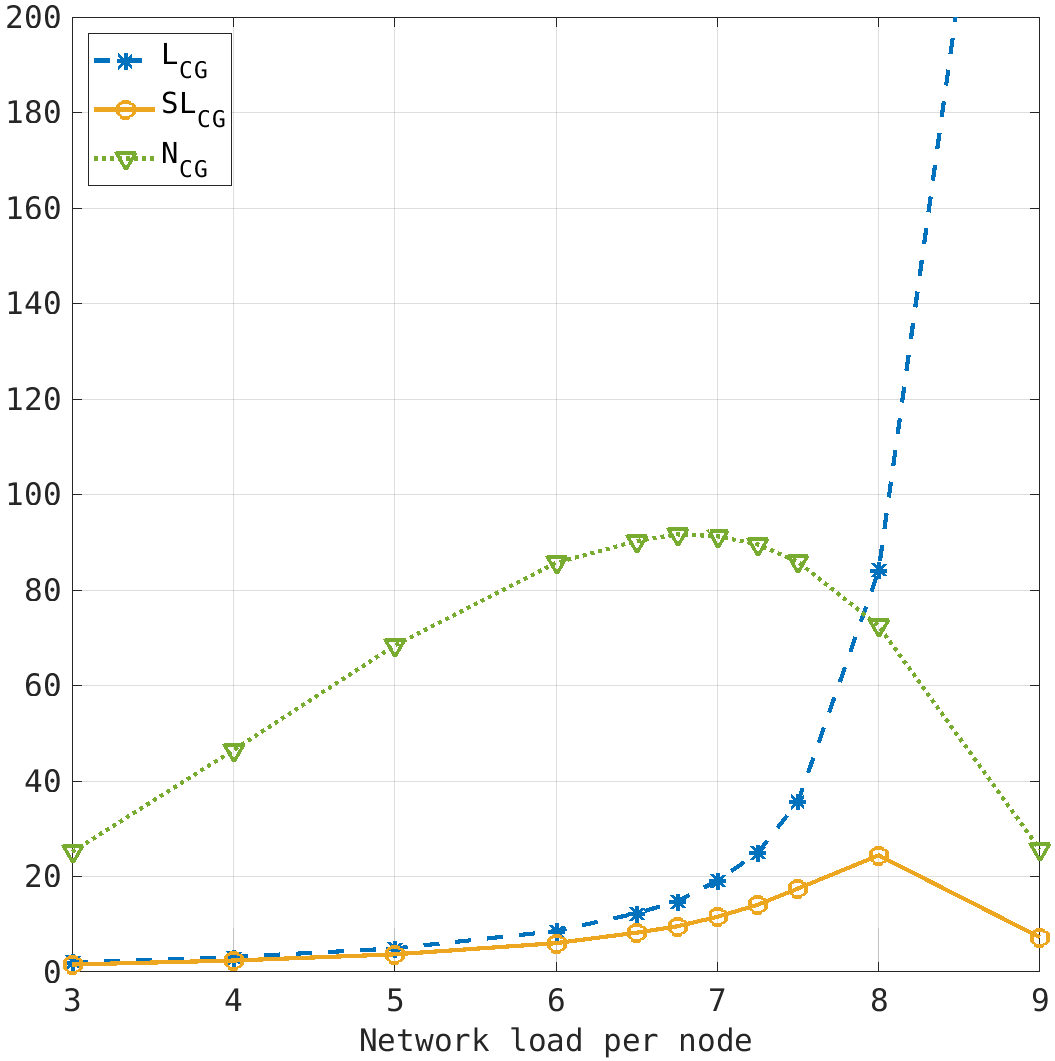}
    \end{minipage}
\hfill
    \begin{minipage}[t]{.49\textwidth}
    \centering
    \includegraphics[width=1\textwidth]{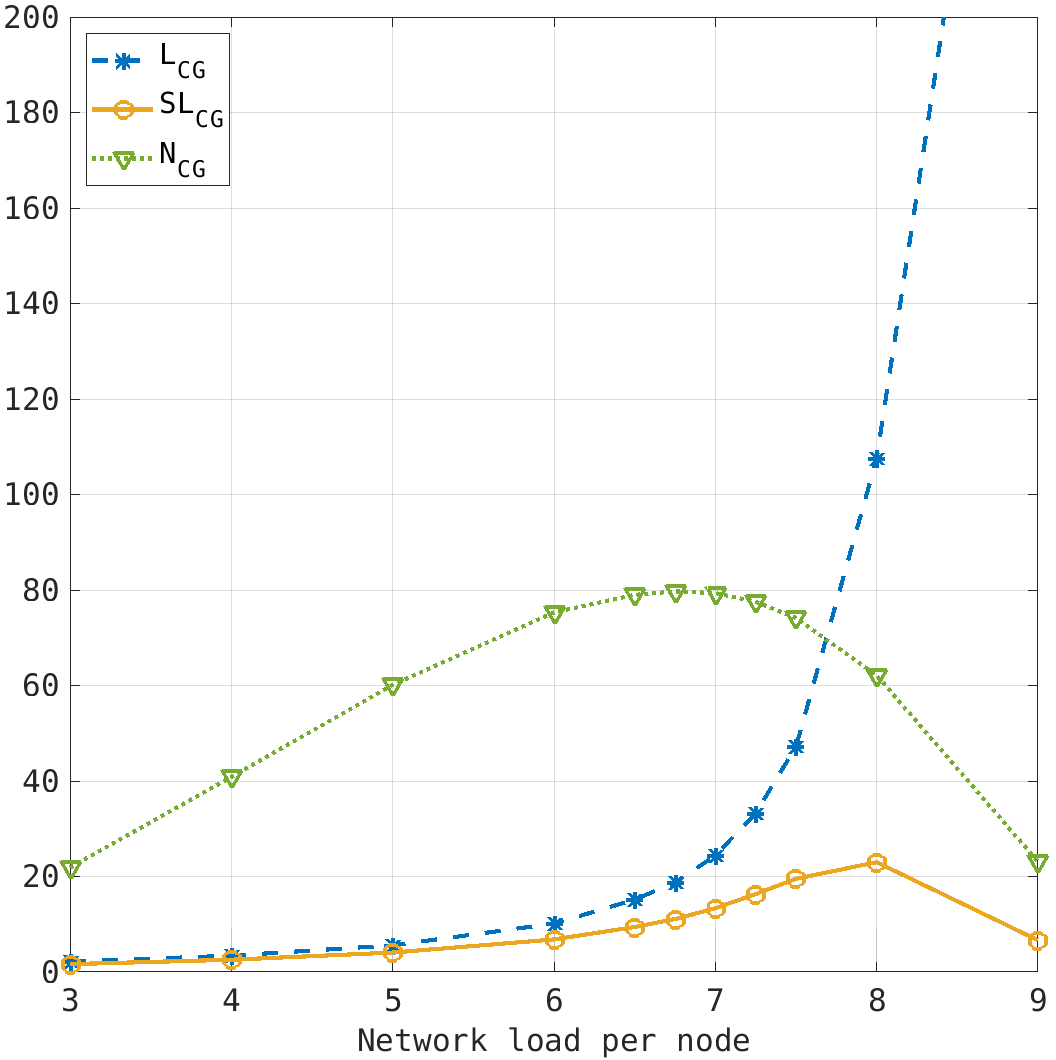}
    \end{minipage}
\caption{Size of the largest congested cluster ($L_{CG}$ curve with blue dots), size of the second-largest congested cluster ($SL_{CG}$ curve with yellow dots) and number of the congested clusters ($N_{CG}$ curve with green dots) as a functions of the average traffic load. The left picture refers to the one step process approximation and the right picture refers to the synchronous dynamics. The percolation transition occurs when a macrocluster of congested nodes emerges in the network ($L_{CG}$ curve) and the numerical results give $\bar n\simeq 8$ for the critical traffic load for both cases (the one step and the synchronous dynamics). At this value the second-larger cluster curve has a maximum whereas the maximum value of the number of congested clusters is achieved at $\bar n\simeq 7$ that suggests as the merging of the small congested clusters anticipate the percolation transition.}
\label{fig:percolation}
\end{figure}

\noindent 
In Fig. \ref{fig:fdiagram} we plot the average flows nodes (\ref{aveflowmax}) with respect to the average traffic load per node $\bar n$ to get the fundamental diagram for the network\cite{daganzo2008} that points out as the maximum flow is reached at $\bar n=n^{\max}/2$ which is lower than the maximum theoretical value since the presence of the congested nodes and empty nodes due to the fluctuations for any traffic load prevents the network transport capacity to reach its maximum value. The fundamental diagram shows that the average flow is almost constant for a large fraction of traffic load $3\le \bar n\le 7$ and it quickly reduces to zero at the limit values $\bar n=0$ and $\bar n=10$. We observe that the traffic load $\bar n=7$ at which the transport capacity begins to drop down, coincides with the critical value at which the clusters of congested nodes start to merge (see Fig. \ref{fig:percolation}). This value can be considered a precursor of a macroscopic congestion formation.   
To explain the fundamental diagram we compute the traffic load distribution $p(n)$ on the network (see Section \ref{congsec} for the definition) as a function of the average traffic load.
The simulation results are shown in Fig. \ref{fig:p_n_3_and_7} where we consider both the one-step process (top pictures) and the synchronous dynamics (bottom pictures). To average on the finite size fluctuations, the empirical stationary distributions are computed applying an ergodic principle and averaging over $10^5$ evolution steps. For the one step process we compare the simulation results with the equilibrium distribution for the single node dynamics (\ref{stazsinglenode}) and we find a perfect agreement between the two approaches. The empirical distributions for $\bar n=3,7$ are also well approximated by the exponential distributions (\ref{expapp0}) and (\ref{expapp1}) as expected by a entropy principle. This is also true in the more realistic case of a synchronous dynamics where we compare the stationary empirical distribution computed by averaging on $10^5$ iterations with the exponential approximation for $\bar n=3,7$. 
The distribution $p(n)$ points out as for any traffic load the probability of observing empty and congested reduces the traffic flow on the network and that over the critical threshold $\bar n=5$ the distribution is peaked at the congested nodes. However the congestion emergence in the network is a continuous process and no singular behavior is observed in the distribution.
In the synchronous evolution we observe that for a low traffic load, the empty state is underexpressed with respect to the exponential interpolation, and that we have states with traffic load $n>n^{max}$. Both these effects are consequence of the degree of the nodes $d\ge 2$: indeed the incoming flows can change the node load up to $d$ particles at the same time so that the empty state frequency is depressed and we have node with a traffic load $>n^{\max}$. But the exponential interpolation is a good approximation of the true stationary distribution according to a local maximum entropy principle. 
To illustrate how the emergence congestion affects the single node distribution we have considered the dependence of the standard deviation on the mean traffic load for the single nodes. The results are reported in Fig. \ref{fig:std_devs_mean} for the average loads $\bar n=3,4,5$. In the cases $\bar n=3,4$ we propose a linear interpolation that would correspond to an exponential like distribution. The point distribution in the pictures reflects the heterogeneity of the node behavior in the transport network that is maximum at the low traffic load $\bar n=3$. This correlation is completely lost at the critical load $\bar n=5$ when the standard deviation is maximum but the heterogeneity of the nodes is minimum. 
However to understand the congestion emergence one has also to consider the spatial distribution of congested nodes to detect the presence of congested clusters.  As previously discussed, the synchronous dynamics introduces a correlation among the states of connected nodes when they are empty or congested. This correlation gives rise to entropic forces that tend to cluster the empty and the congested nodes. To highlight this effect we compute the stationary distribution when $\bar n=n^{\max}/2$. The results for the considered random network are shown in Fig. \ref{fig:p_n_5} both for the single step process and the synchronous dynamics. We observe that in the first case the distribution is almost flat, whereas in the second case the presence of the entropic forces induce a bimodal distribution with peaks at the empty and congested nodes. 
    

\begin{figure}
    \begin{minipage}[t]{.5\textwidth} 
    \centering
    \includegraphics[width=1\textwidth]{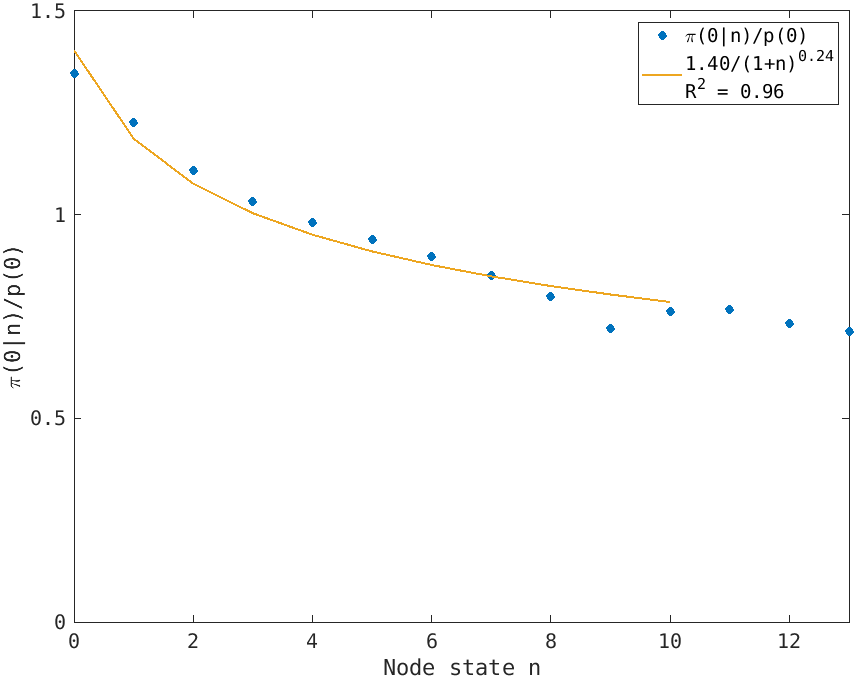}
    \end{minipage}
\hfill
    \begin{minipage}[t]{.5\textwidth}
    \centering
    \includegraphics[width=1\textwidth]{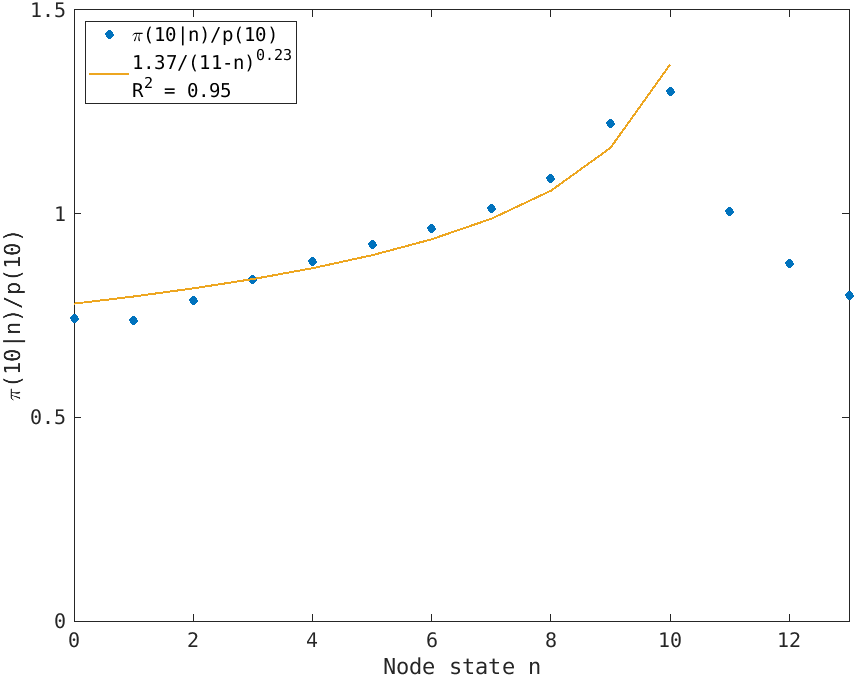}
    \end{minipage}
\caption{Left picture: the normalized conditional probability $\pi(0\vert n)/p(0)$ for the transport network model is numerically computed using the synchronous dynamics(dots). The average traffic load is $\bar n=5$. The continuous line is a proposed interpolation with the power law function reported in the inset together with the corresponding $R^2$-value.
Right picture: the same as in the left picture for the conditional probability $\pi(n^{\max}\vert n)/p(n^{\max})$(dots). The continuous line corresponds to the power law interpolation reported in the inset.}
\label{fig:synchro_cond_p_5}
\end{figure}

\begin{figure}
    \begin{minipage}[t]{.5\textwidth} 
    \centering
    \includegraphics[width=1\textwidth]{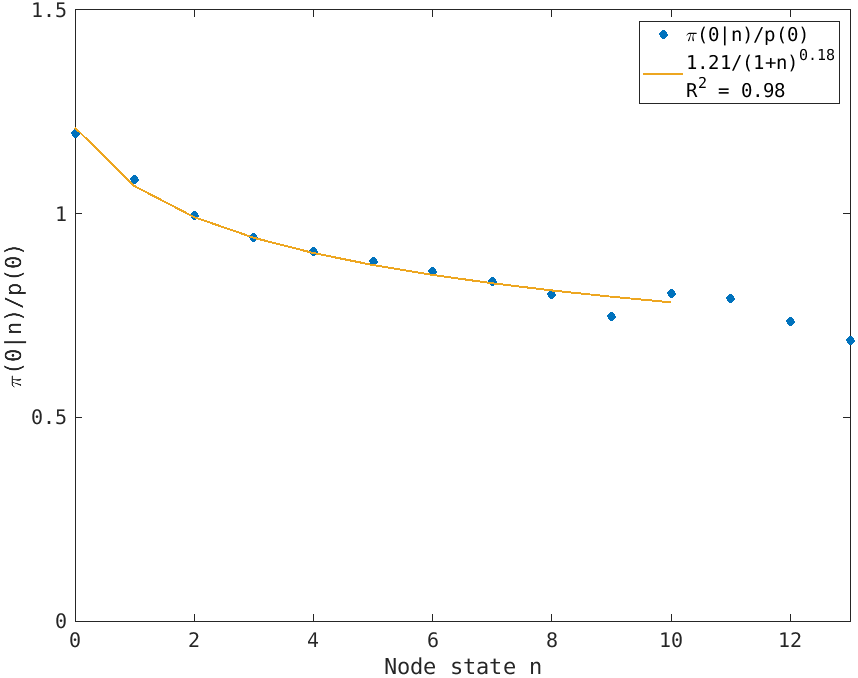}
    \end{minipage}
\hfill
    \begin{minipage}[t]{.5\textwidth}
    \centering
    \includegraphics[width=1\textwidth]{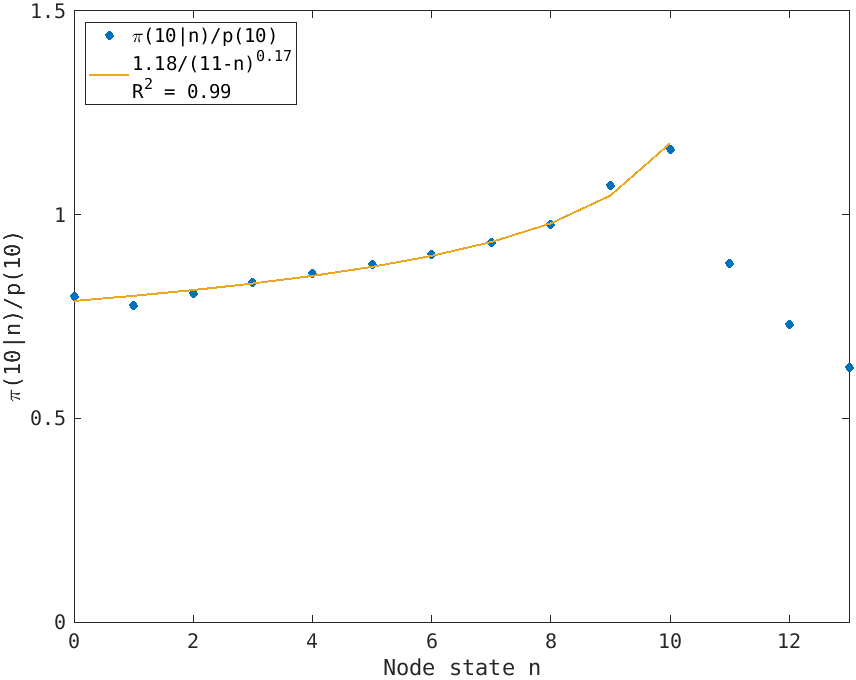}
    \end{minipage}
\caption{Left picture; the normalized conditional probability $\pi(0\vert n)/p(0)$(dots) is computed (cfr. Fig. \ref{fig:synchro_cond_p_5} left) using an average traffic load $\bar n=3$. In the inset it is reported the power law interpolation (continuous line).
Right picture: the normalized conditional probability $\pi(n^{\max}\vert n)/p(n^{\max})$(dots) is computed (cfr. Fig. \ref{fig:synchro_cond_p_5} right) using an average traffic load $\bar n=7$. In the inset it is reported the power law interpolation (continuous line).}
\label{fig:synchro_cond_p_3_and_7}
\end{figure}

\begin{figure}
    \begin{minipage}[t]{.5\textwidth} 
    \centering
    \includegraphics[width=1\textwidth]{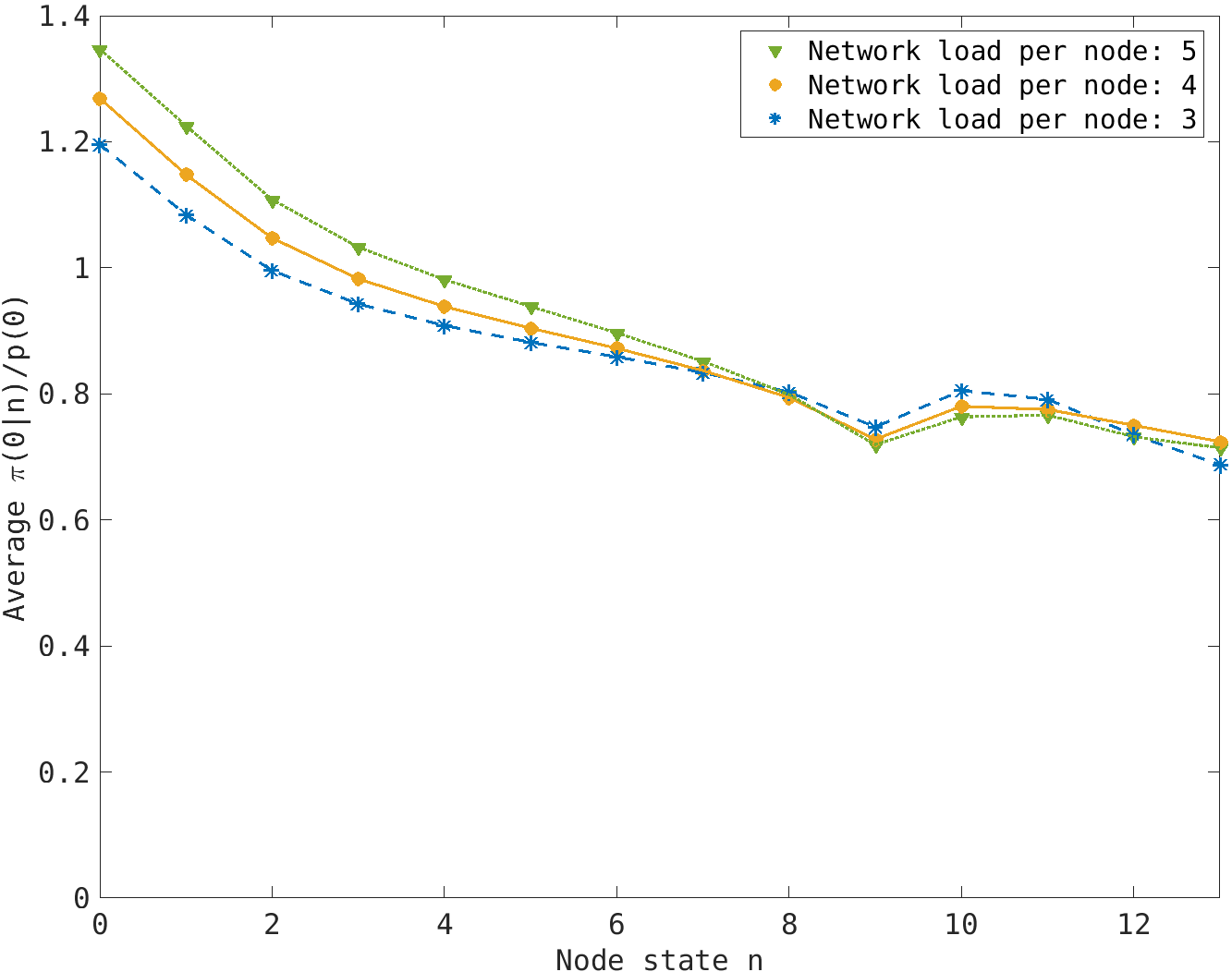}
    \end{minipage}
\hfill
    \begin{minipage}[t]{.5\textwidth}
    \centering
    \includegraphics[width=1\textwidth]{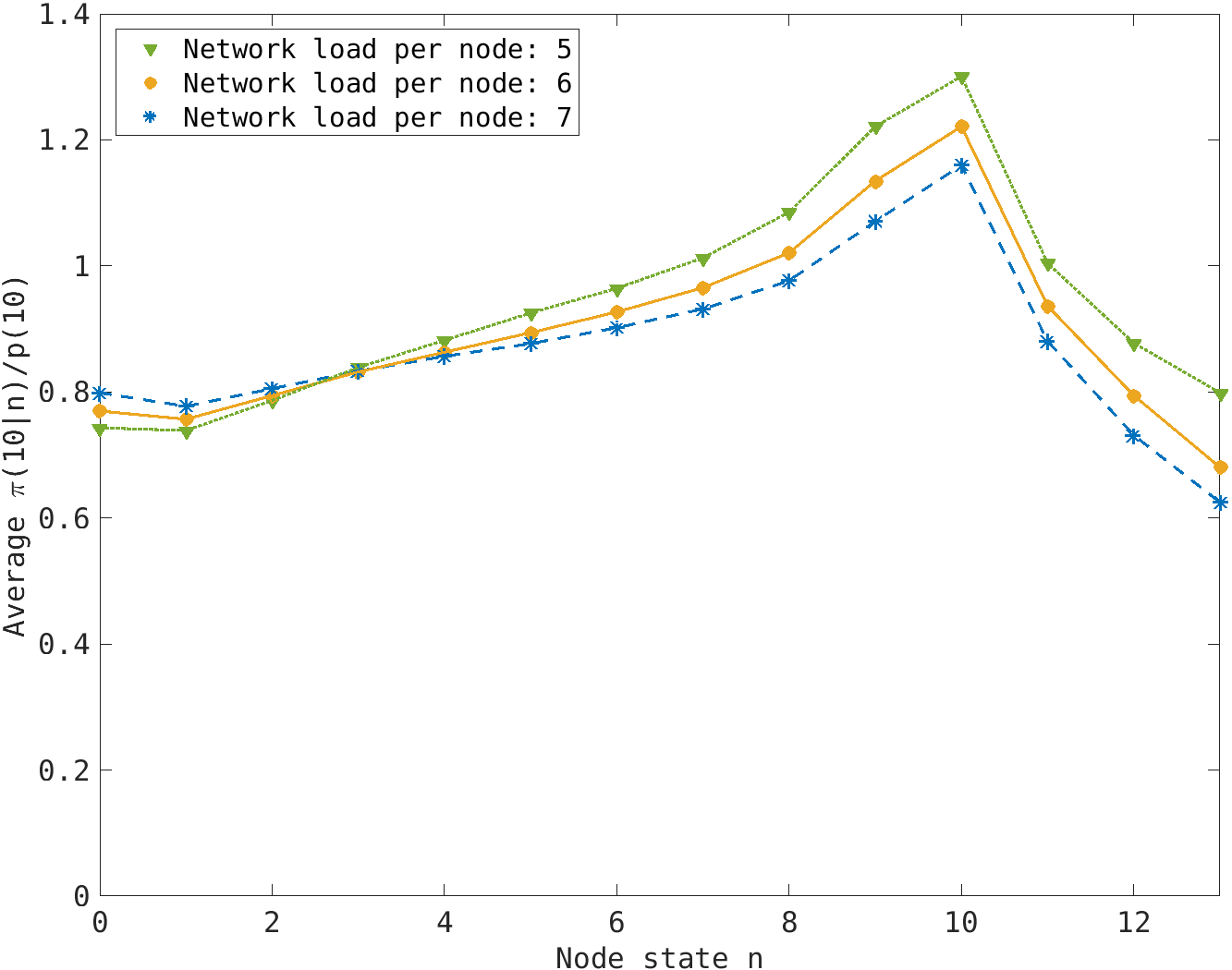}
    \end{minipage}
\caption{Left picture: normalized conditional probability $\pi(0\vert n)/p(0)$ for the transport network model using the synchronous dynamics for the average traffic loads $\bar n=3,4,5$; different symbols are used to distinguish the cases as reported in the inset. Right picture: the same as in the left picture for the normalized conditional probability $\pi(10\vert n)/p(10)$ with average traffic loads $\bar n=7,6,5$.}
\label{fig:synchro_cond_p_all}
\end{figure}

To quantify the entropic forces we have computed the conditional probabilities $\pi(0|n)$ and $\pi(n^{\max}|n)$ that denote the probability that a node with load $n$ has a empty neighbor or a congested neighbor respectively. In the case of the single step process these probabilities are 
$$
\pi(0|n)\simeq p(0)\qquad \pi(n^{\max}|n)\simeq p(n^{\max})
$$
since the nodes evolve independently in the thermodynamic limit.
Conversely, in the synchronous case we expect that the correlation among the node states will increase the values $\pi(0|0)$ and $\pi(n^{\max}|n^{\max})$ near the boundary states $n=0$ and $n=n^{\max}$. The numerical results are reported in Fig. \ref{fig:synchro_cond_p_5} and \ref{fig:synchro_cond_p_3_and_7} where the effect of the entropic force is clearly visible at low and high traffic load and an interpolation of the numerical results by a power law distribution is proposed.
\begin{align}
\label{interp}
    \pi(0|n)&\propto\frac{p(0)}{(1+n)^\alpha}\nonumber \\
    \pi(n^{\max}|n)&\propto \frac{p(n^{\max}}{(n^{\max}+1-n)^\alpha}
\end{align}
In the numerical simulations we have initially considered  an average traffic load $\bar n=n^{\max}/2$ (Fig. \ref{fig:synchro_cond_p_5}) where the interpolation (\ref{interp}) of the numerical results highlights the symmetry of the entropic forces when we consider the effect on the empty and the congested nodes. Then we have simulated a low traffic load $\bar n_l\ll n^{\max}$ and a high traffic load $\bar n_h\simeq n^{\max}$ but with the symmetric condition $\bar n_h=n^{\max}-\bar n_l$ (Fig. \ref{fig:synchro_cond_p_3_and_7}). The interpolation (\ref{interp}) provides almost the same exponent in the two cases consistently with the equivalence of the gap dynamics and the particles dynamics in the case of a balanced network.
The case $\bar n=n^{\max}/2$ is critical since the exponent is maximal (so one has a faster decaying of the correlation) suggesting that the effect of the two entropic forces tends to balance when the traffic load of the nodes is far from the limit values. This effect is illustrated by Fig. \ref{fig:synchro_cond_p_all} where we compute the normalized conditional probabilities $\pi(0|n)|p(0)$ and $\pi(n^{\max}|n)/p(n^{\max})$ for the critical traffic load $\bar n=5$. The results also point out the symmetrical behavior of the particle and gap dynamics (except near the boundary values $n=0$ and $n\ge n^{\max}$).


To see how the entropic forces depend on the traffic load we have computed the the normalized conditional probabilities $\pi(0\vert n)/p(0)$ and  $\pi(n^{\max}\vert n)/p(n^{\max})$ for a traffic load $\bar n=3$ (see Fig. \ref{fig:synchro_cond_p_3_and_7} left) and $\bar n=7$ (see Fig. \ref{fig:synchro_cond_p_3_and_7} right) respectively. The results suggest that the maximum effect of the entropic forces is achieved at the critical load $\bar n=n^{\max}/2$ whereas the connected node states tend to be weakly correlated at low or high traffic loads. This can be understood since the probability $p(0)$ increases for low traffic load decreasing the effect of the entropic force and similarly it happens to $p(n^{\max}$ at high traffic load. This explains the quite good exponential interpolation of the distribution $p(n)$ for the synchronous case (see Fig. \ref{fig:p_n_3_and_7}) that is computed maximizing the entropy (i.e. considering the nodes independent in the thermodynamic limit). 
To illustrate this effect we compare the normalized conditional probabilities $\pi(0\vert n)/p(0)$ and  $\pi(n^{\max}\vert n)/p(n^{\max})$ for traffic loads $\bar n=3,4,5$ and $\bar n=7,6,5$ respectively. The results are reported in Fig. \ref{fig:synchro_cond_p_all} where we observe as the entropic forces have a maximum effect at $\bar n=5$ and the symmetry of the particles and gaps dynamics for the homogeneous transport network. 
According to the previous results the traffic load $\bar n=n^{\max}/2$ is a critical value for the system fluctuations, but the emergence of a macroscopic congestion depends on the distribution of the congested nodes in the transport network when the traffic load increases. The formation of a giant cluster is the fingerprint of the percolation transition that it has been proposed to describe the congestion on an urban road network\cite{ambuhl2023}. We have computed the number of congested clusters, the largest cluster size and the second largest cluster size for the transport network model considering different traffic load, both in the case of the one step process approximation and in the synchronous dynamics. The simulation results are shown in Fig. \ref{fig:percolation} where the percolation transition occurs when the largest cluster size steepens whereas both the number of clusters and the second largest cluster size decrease due to the formation of a large congested cluster. The results give a percolation threshold when the average traffic load is $7< \bar n <8$ both for the one step process and the synchronous dynamics: the main difference is observed in the number of congested clusters which is lower in the case of synchronous dynamics. This is consistent with the presence of the entropic forces that tends to cluster the congested nodes (so that their number reduces) and also the dimension of the largest cluster is bigger than in the synchronous case. The coincidence of the percolation threshold in both cases can be explained by the weak effect of the entropic forces at high traffic load. The formation of a giant congested cluster can destroys the connectivity of the transport network (percolation transition) even if the cluster is not static in model since the outgoing flow from the congested nodes moves the cluster in the network.
However the emergence of percolation transition requires

\section{Conclusions}

We adopted a reductionist approach to highlight the universal properties of the congestion transition in a urban road network. Our approach uses a simple transport network represented by a Markov random process on a graph, where nodes symbolize specific locations, and weighted links dictate the transition rates between these nodes. By setting a finite transport capacity and a maximum capacity at each node, we were able to simulate the typical feature of a urban traffic regulated by the crossing point. Two different dynamics are considered: the one step process where a randomly chosen node moves each time step and the synchronous dynamics where all the nodes move together using the same information on the network state. Our main goal is to explain the role of traffic fluctuations on the congestion transition and characterize the types of dynamics. 
Throughout our analysis, we maintained the assumption that the transport network is in a balance condition (cfr. def. (\ref{balance1})), where the average incoming flow at each node equals the average outgoing flow, reflecting the realization Wardrop equilibrium where paths are optimized across the network to do not create congestion. In the case of non balanced transport networks there exist nodes that are certainly congested when the traffic load increases and that play the role of hot spots for the congestion formation. Conversely, if the balance condition is satisfied the congestion is not driven by the average dynamics, but by the fluctuations. Our analytical methods, particularly the application of a Maximum Entropy Principle, allowed us to characterize the distribution of traffic load fluctuations across nodes and identify how congestion emerges in a continuous way as the overall traffic load increases. Notably, we study the entropic forces in synchronous dynamics, leading to the clustering effect of congested and empty nodes.
\par\noindent
The stationary distribution of traffic load fluctuations can be approximated by a local entropy principle and the congestion transition
is characterized by a peak of traffic load variance of the nodes, yet without singularity in the thermodynamic limit. Numerical simulations further demonstrated the formation of a macroscopic congested cluster, resembling a percolation-like transition within the network, requiring traffic loads nearing maximum capacity for its emergence.
The simulations highlight the effect of the entropic forces in synchronous process in the congestion transition as a greater dimension of the congested clusters with respect the one step process, but these slightly affect the value of critical traffic load for the percolation transition. 
\par\noindent
Our model is certainly a simplification of the microscopic urban traffic complexity but it allows to underscore some universal macroscopic features. Our results explain the emergence of congestion due as a continuous processes that creates clusters of congested nodes mimicking the percolation transition (i.e. the formation of a macroscopic congested cluster at high traffic load). The existence of entropic forces depends on the network dynamics and favors the cluster formation of empty and congested nodes. This research not only provides insights into the nature of traffic congestion but also contributes to the theoretical framework necessary for enhancing urban transportation systems. The application of control strategies to mitigate the congestion effect requires two further steps. On one hand, the time scales of congestion formation related to the relaxation time scales of the system, have to be studied to estimate the congestion diffusion in the transport network, On the other hand, one has to develop strategies to reduce the fluctuation variance of the local traffic load across the whole network using the crossing dynamics.

\section {Acknowledgments}
This study was carried out within the MOST - Sustainable Mobility National Research Center and received funding from the European Union Next-GenerationEU (PIANO NAZIONALE DI RIPRESA E RESILIENZA (PNRR) - MISSIONE 4 COMPONENTE 2, INVESTIMENTO 1.4 - D.D. 1033 17/06/2022, CN00000023).



\appendix
\section{Analysis of the stationary distributions}
\label{append1}
Using a dynamical point of view the stationary distribution is proportional to the visiting frequency of any microstate during the evolution when the ergodic property a holds.

The normalizing constant in the eq. (\ref{distri1}) $C_N^{n^{\max}}(\vec p)$ can be defined in a recursive way using a generating function 
\begin{equation}
 C_N^{n^{\max}}(\vec p)=\sum_{|n|=N}^{n^{\max}}\prod_{i=1}^M p_i^{n_i}\simeq \frac{1}{N!}\left.\frac{d^N}{dx^N}\right \vert_{x=0}
\prod_{i=1}^M \frac{1}{(1-p_i x)}
\label{cncomp}
\end{equation}
in the limit $n^{\max}\gg 1$. An analytical estimate of the normalizing constant is obtained using the Cauchy theorem
$$
C_N(\vec p)\le \prod_{i=1}^M \frac{1}{1-p_i}
$$
The function 
$$
g(x)= \prod_{i=1}^M \frac{1}{1-p_ix}
$$
is analytic in the disk $|x|\le 1$ and it has poles at $x=p_i^{-1}$, so that the radius of convergence $R_\ast$ of the Taylor expansion at $x=0$ depends on the nearest pole to the origin: $R_\ast=\min_{i=1,...,M} (p_i^{-1})$. We recover that if $p_i$ are not uniform $C_N(\vec p)$ is dominated by the contribution of the greatest $p_i$. \par\noindent 
Let $p_i=M^{-1}$, then all the microstates have the same weight and we have a simple solution for the partition function $C_N(p)$ that corresponds to the number $C_N(M)$ of $M$-components vectors $\vec n$ with non negative integer vectors such that $|\vec n|=N$
\begin{equation}
C_N(M)=\frac{1}{M^N}\binom{N+M-1}{M-1}
\end{equation}
The stationary distribution becomes
\begin{equation}
\rho_s(\vec n)=\binom{N+M-1}{M-1}^{-1}
\label{stazsol2}
\end{equation}
and it maximizes the Gibbs Entropy (\ref{gibbsent}) by definition. Since this is an attractive distribution we expect that the one-step process (\ref{master1}) increases the entropy during the evolution. 
We remark that for $N$ finite the state of the nodes are not independent since the state of each node affects the other node states, but they can be considered independent for $N\to\infty$. The empty state is a boundary condition state at which the outgoing flow from the node is zero. If one chooses randomly the network state $\vec n$, the probability that a node has $n$ particles requires to estimate the mean value of
$$
\frac{\sharp\{|\vec n|=N, n_k=n,k=1,...,M\}}{M}=\frac{1}{M}\sum_k \delta_{n,n_k}
$$
Then one can compute the average value 
$$
E(\delta_{n,n_k})=\sum_{|\vec n|=N, n_k=n}\rho_s(\vec n)
$$
In the balanced case we have
$$
\sum_{|\vec n|=N, n_k=n}\rho_s(\vec n)=\frac{C_{N-n}(M-1)}{C_N(M)}
$$
It follows
$$
E(\delta_{n,n_k})=\binom{N-n+M-2}{M-2}\left [\binom{N+M-1}{M-1}\right]^{-1}=
\frac{(M-1)N...(N-n+1)}{(N+M-1)...(N-n+M-1)}
$$
and for $N,M\gg 1$
$$
\frac{\sharp \{|\vec n|=N, n_k=n\}}{M}\simeq \frac{1}{1+\bar n}\frac{1}{(1+\bar n^{-1})^n}
$$
A direct calculation shows
$$
\frac{1}{1+\bar n}\sum_{n\ge 0}\frac{1}{(1+\bar n^{-1})^n}=\frac{1}{1+\bar n}\frac{1+\bar n^{-1}}{\bar n^{-1}}=1
$$
and the average value is
$$
E(n)=\frac{1+\bar n^{-1}}{1+\bar n}\sum_{n\ge 1}\frac{n}{(1+\bar n^{-1})^{n+1}}=
-\frac{1}{\bar n}\frac{d}{d\bar n^{-1}}\sum_{n\ge 0}\frac{1}{(1+\bar n^{-1})^n}
$$
and we recover
$$
E(n)=-\frac{1}{\bar n}\frac{d}{d\bar n^{-1}}(1+\bar n)=\frac{1}{\bar n^{-1}}=\bar n
$$
where the nodes are all equivalent. In the thermodynamic limit the marginal distribution for a single node $p_k(n)$ is
\begin{equation}
p_k(n)=\frac{(1+\bar n)^{-1}}{(1+\bar n^{-1})^n}\simeq \frac{1}{\bar n}e^{-n/\bar n}
\label{limit}
\end{equation}
where the last estimate requires $\bar n$ not too small. Since the nodes are equivalent the marginal distribution $p_k(n)=p(n)$ gives the probability to observe the state $n$ for any node and it describes the distribution of the node states in a large network $N\gg 1$. This distribution is associated to the fluctuations of the node state with respect the mean value $\bar n$ (but the time scale of the fluctuations depend on the spectral properties of the Laplacian matrix of the master equation (\ref{master1}). Since the stationary distribution of the network state maximizes the entropy, the same is expected for the marginal distribution with the constraint on the state of the $k$-node and the total number of particle. Indeed in the thermodynamic limit we have  the maximal entropy solution with the constraint $E(n)=\bar n$ for an ensemble of equiprobable microstates and the state of the $k$-node becomes independent from the states of all the other nodes. We remark that this result holds for any transport network in a balance condition, whatever is the structure of the network. In such a case the dynamics is dominated by the fluctuations that scale as $\bar n$ (not  $\sqrt{\bar n}$ as for the Poisson distribution). 
The nodes with a population below the average value are over-expressed and few nodes have a great population, but the fluctuations continuously change the state of each node redistributing the particles.\par\noindent 
The Markov process associated to the master equation (\ref{master1}) is realized by choosing randomly a link $j\to i$ in the graph and, if $n_j>0$ moves a particle with probability $\pi_{ij}\Delta t$. Then network state changes and a new movement is performed. 
\par\noindent
The counterpart of the one step process is the synchronous dynamics when at each time step all the nodes can move one particle if their state is not empty. For a given
state $\vec n$ the outgoing flows are
$$
\Phi^-(\vec n)=\sum_{j_1,...,j_M=1}^M \pi_{j_11}(n_1)...\pi_{j_MM}(n_M)\rho(\vec n,t)
$$
where one considers all the possible exchanges among the nodes so that many of the terms are zero
in the previous sum. However according to the definition of the transition rates $\pi_{ij}(n)$ it is necessary 
to extend the definition by setting $\pi_{ii}(0)=1$ otherwise the product which defines the probability rate vanishes each time a configuration contains a null node.  In a similar way, the incoming flow can be written in the form
\begin{align}
\label{flowsync}
&\Phi^+(\vec n)=\sum_{j_1,...,j_M=1}^M E_{j_1}^-...E_{j_M}^-  E_1^+... E_M^+ \pi_{j_1 1}(n_1) ... \pi_{M j_M}(n_M)\rho(\vec n,t)\nonumber \\
&= \sum_{j_1,...,j_M=1}^M  \pi_{j_1 1}(n_{1}-m_1+1) ... \pi_{j_M M}(n_{M}-m_{M}+1) \rho\left (\vec n+\vec 1-\sum_k \hat e_{j_k},t \right )
\end{align}
where $m_k$ counts the repetitions number of the index $k$ in the $M$-nuple $j_1,..,j_M$ (i.e. $m_k$ defines the  initial configuration $n_k-m_k+1$ of the $k$ node so that it receives $m_k$ particles to get the final configuration $n_k$). We remark that in the synchronous dynamics all the nodes move  referring to the same network state, since the state updates after all the displacements.\par\noindent
In a balance condition one considers all the possible exchanges with the connected nodes assuming that all the nodes are not empty and using the transition probability rates one has $<m_k>=1$ (the average incoming flow equals the outgoing flow).  If a node is connected to empty nodes the incoming flow reduces (in average) whereas the outgoing flow is fixed. This means that there exists an effective 'force', that attracts the node state to the zero state when it is connected to empty nodes. Therefore a node gets to a neutral equilibrium state when it is connected to non empty nodes and the balance condition holds, but each time one of its neighborhoods is empty there is a negative average net flux. The converse is also true, if we consider a node in a zero state: the net average flux is always positive and it is maximal when all the connected nodes are not empty. 
This effect gives an explanation to the exponential limit distribution as the solution of a diffusion process with a boundary conditions and there exists a constant force toward the boundary. We analyze the node dynamics near the zero state in detail, but a similar argument can apply to the nodes near the congested state. If $n^{\max}\gg 1$ the stationary condition for the flows at a fixed node is
$$
(1-p_0)\sum_j \pi_{ij}=\sum_j \pi_{ji}(1-p_0)
$$
that it is always satisfied in a balance condition when the probability of the zero state $p_0$ is the same for all the nodes. 
The average number of zero states in the stationary distribution is a property of the dynamics. In the one step process dynamics all the configurations are equiprobable and to compute the probability to observe a certain number of zero states one has to count the number of possible configurations distributing of the zero states in the network. In the thermodynamic limit the fraction of zero states can be computed by the distribution (\ref{limit}) and the nodes can be considered independent. In the synchronous dynamics the presence of the zero states reduces the expected incoming flows to the boundary nodes: i.e. the nodes connected to the empty nodes. A node $k_b$ is a boundary node of a given network state $\vec n$ ($k_b\in B(\vec n)$) if there exists $j$ s.t. $\pi_{k_bj}\ne 0$ with $n_j=0$. Let $M_0$ the number of empty nodes of the state $\vec n$
$$
M_0=\sharp\{j: |\vec n|=N,\quad n_j=0\quad j=1,...,M \},
$$
the number of boundary nodes depends on the distribution of the zero states in the network. The number of the boundary nodes is the volume $V=\sharp B(\vec n)$. It is clear that the volume $V_0(\vec n)$ is minimal when the zero state nodes form a cluster. In the evolution of the transport network model, the nodes at the boundary feel an effective force that decreases the their state toward zero. If $V_0(\vec n)$ is minimal the total`force' acting on the node states at a given time is minimal and this corresponds to an `equilibrium situation' so that we expect that these configurations are favored in a the stationary condition. Under this point of view we call these forces `entropic' since they change the microstate probability favoring the state with minimal boundary volume $V_0(\vec n)$. This effect is counteracted by the fluctuations of the due to the stochastic dynamics that increases the configuration entropy distributing the particles among the nodes, so that the effect of entropic force decreases when the node degree increases. Since the entropic forces in the synchronous dynamics decrease the total entropy of the stationary distribution, the non-reversibility character implies that one has a non-equilibrium stationary state with an entropy production. An open question is if the stationary state can be characterized by a minimal entropy production\cite{prigo1}.\par\noindent
The effect of the entropic forces on a node depends on its connectivity: a node with high degree is little affected by having an empty neighbor. Therefore we expect a clustering effect of zero state nodes that have a low incoming degree. In a highly connected graph the entropic forces are negligible and one can approximate the invariant measure by a uniform measure as in the one step process.\par\noindent 
For a given number of zero state nodes the configuration with a minimal boundary volume depends on the graph structure: in principle this could be related to the existence of a community that are poorly connected with the rest of the network. This problem requires to define a geometry for the network to understand how empty node clusters appear and increase in the transport network. 
\end{document}